\documentclass{scrartcl}
\usepackage[numbers,sort&compress]{natbib}
\usepackage[margin=1in]{geometry}
\usepackage{bm,epsfig,enumerate,tabularx,lmodern,fix-cm,mathtools,xcolor}

\newcommand{\etal}{\textit{et al}.}

\begin{document}

\title{Assessing the robustness of spatial pattern sequences in a dryland vegetation model}

\author{Karna Gowda\thanks{Department of Engineering Sciences and Applied Mathematics, Northwestern University, Evanston, IL 60208, USA}, Yuxin Chen\footnotemark[1], Sarah Iams\thanks{Paulson School of Engineering and Applied Sciences, Harvard University, Cambridge, MA 02138, USA}, Mary Silber\thanks{Department of Statistics, The University of Chicago, Chicago, IL 60637, USA}}
\date{}

\maketitle

\begin{abstract}
A particular sequence of patterns, ``$\text{gaps} \to \text{labyrinth} \to \text{spots}$,'' occurs with decreasing precipitation in previously reported numerical simulations of PDE dryland vegetation models. These observations have led to the suggestion that this sequence of patterns can serve as an early indicator of desertification in some ecosystems. Since parameter values can take on a range of plausible values in the vegetation models, it is important to investigate whether the pattern sequence prediction is robust to variation. For a particular model, we find that a quantity calculated via bifurcation-theoretic analysis appears to serve as a proxy for the pattern sequences that occur in numerical simulations across a range of parameter values. We find in further analysis that the quantity takes on values consistent with the standard sequence in an ecologically relevant limit of the model parameter values. This suggests that the standard sequence is a robust prediction of the model, and we conclude by proposing a methodology for assessing the robustness of the standard sequence in other models and formulations.
\end{abstract}

\section{Introduction} 
\begin{figure}[!t]
	\centering \includegraphics[width=0.52\columnwidth]{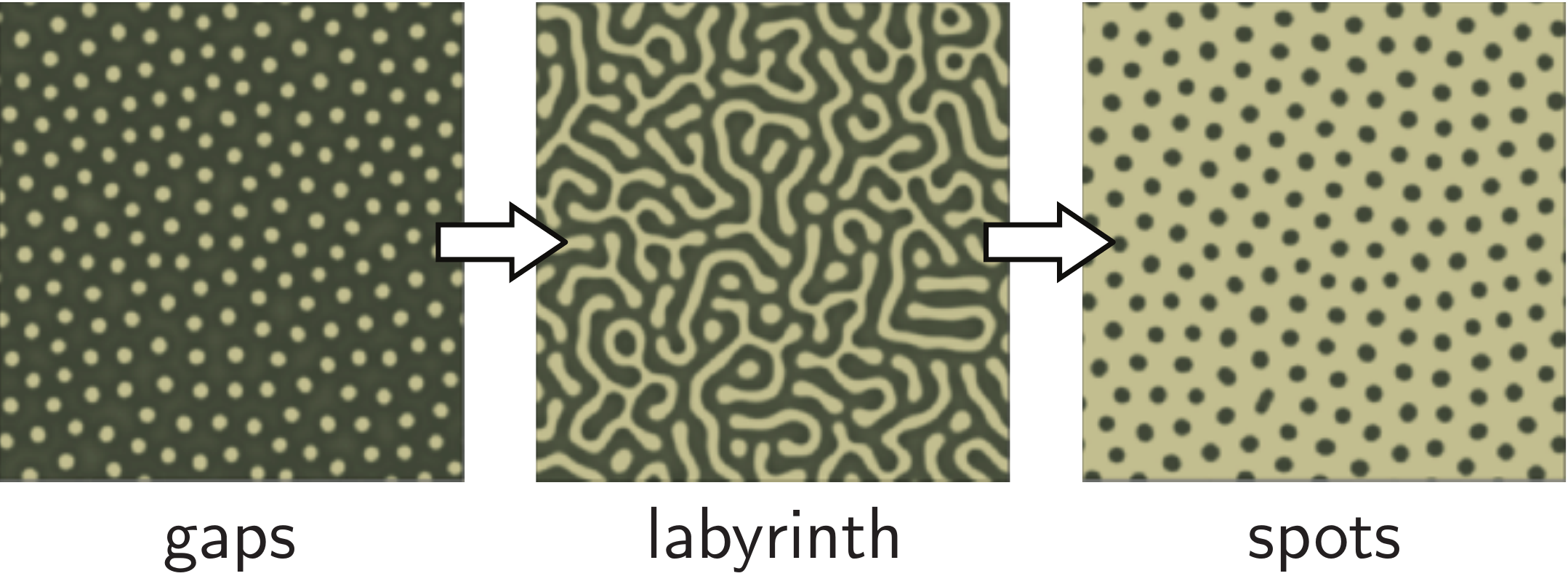} \caption{\small Example of the standard ``gaps $\to$ labyrinth $\to$ spots'' sequence in the vegetation model by Rietkerk \etal. Qualitatively different patterns occur at successively smaller values of a precipitation parameter. Darker shading denotes higher levels of vegetation biomass.} \label{fig:sequence_example}
\end{figure}

Many studies of spatially periodic patterns in models of dryland vegetation focus on patterns as potential indicators of ecosystem vitality~\citep{vonHardenberg:2001bka, Bel:2012jqa,Dakos:2011dpa,Gilad:2004bp,Gowda:2014cd,Guttal:2007ixa,Kefi:2010gca,LeJeune:2002eoa,Meron:2004dra,Sherratt:2015cq,Zelnik:2013iv,vanderStelt:2013wy,Siteur:2014jm}. In particular, flat-terrain patterned states in several models have been shown in simulations to evolve through a sequence of morphologies, ``$\text{gaps} \to \text{labyrinth} \to \text{spots}$'' (Figure \ref{fig:sequence_example}), as ecosystem aridity increases~\citep{vonHardenberg:2001bka, Rietkerk:2002ufa, LeJeune:2004bma, Gilad:2004bp}. This sequence precedes the collapse of vegetation in the models, which has led to the suggestion that real ecosystems may evolve through this predictable sequence of patterns en route to desertification~\citep{Rietkerk:2004vqa, Scheffer:2009wj}. In this way, vegetation patterns may serve as early-warning signs of critical ecosystem transitions. 

The pattern sequence prediction emerges from a modelling framework comprising a number of different ecological hypotheses, functional formulations, and restrictions on plausible parameter sets. It is therefore important to investigate whether the prediction is robust within this framework. It is encouraging that multiple models can produce the pattern sequence ``$\text{gaps} \to \text{labyrinth} \to \text{spots}$'' (the standard sequence), but little work has been done to assess whether the sequence occurs in any model under variations of the parameter values, which may vary significantly between ecosystems. In this paper, we ask whether easily calculable quantities, derived from bifurcation theoretic analysis of pattern-forming instabilities, can be used to predict where the standard sequence occurs in a model's parameter space. To do this, we analyse and simulate the widely studied model by Rietkerk \etal~\citep{Rietkerk:2002ufa} across a broad range of parameter values.

The standard sequence, ``$\text{gaps} \to \text{labyrinth} \to \text{spots}$'', is observed in a suite of partial differential equation (PDE) models that describe the community-scale dynamics of dryland vegetation over flat terrain~\citep{vonHardenberg:2001bka, Rietkerk:2002ufa, Gilad:2004bp, LeJeune:2004bma}. These models all involve a Turing instability~\citep{Turing:1952vn} in the formation of patterns from uniform vegetation. Observations of the standard sequence as a model prediction come primarily from numerical simulations~\citep{vonHardenberg:2001bka, Rietkerk:2002ufa, Gilad:2004bp}, which show gap, labyrinth, and spot patterns occurring at successively lower values of a precipitation parameter (Figure \ref{fig:sequence_example}). A study by LeJeune \etal~\citep{LeJeune:2004bma} used bifurcation analysis to demonstrate analytically that this sequence occurs in a tractable 1-field model for a particular parameter set. The apparent agreement between these observations, which come from different model formulations, provides some support for the standard sequence as a robust prediction of this suite of models.

Empirical support for this sequence comes chiefly from two studies of remotely-sensed imagery. A 2006 study by Barbier \etal~\citep{Barbier:2006jwa} used imagery over southwest Niger to demonstrate that gap patterns emerged from uniform vegetation over a period of time coinciding with a prolonged Sahelian drought. This result is consistent with model observations of gap patterns occurring near the onset of pattern formation. A 2011 study by Deblauwe \etal~\citep{Deblauwe:2011ee} classified pattern morphologies in imagery over Sudan and found that different morphologies vary over spatial precipitation gradients in accordance with the standard sequence prediction. Gaps tended to occur in areas with relatively high mean annual precipitation, spots occurred in areas with relatively low precipitation, and labyrinths occurred in between. Pattern dynamics were also assessed using three sets of images taken over a 35-year span. Gaps in some areas were shown to transition to labyrinths over a period of time again coinciding with a sustained regional drought. Labyrinths transitioned to spots in different areas over the same period of time. Though neither of these studies show the standard sequence preceding the collapse of vegetation, they demonstrate consistency between model predictions and empirical observations.

Since these empirical studies focus on two Sub-Saharan regions, they make statements about particular ecosystems, which are subject to particular environmental and ecological parameters. It is unknown how other periodically-patterned flat-terrain ecosystems behave in response to changes in aridity. It is also unknown whether the suite of dryland vegetation models predicts the standard sequence to occur under most circumstances, since the numerical simulations underlying the standard sequence model observations were generated using a small number of parameter sets. The parameter sets appropriate for different ecosystems can differ significantly. For instance, the model for banded vegetation patterns on slopes by Klausmeier~\citep{Klausmeier:1999woa} distinguishes between shrublands and grasslands, using plant mortality parameters that differ by an order of magnitude for these different plant types\footnote{Klausmeier~\citep{Klausmeier:1999woa} uses the mortality rate $M_{tree}=0.18$ year$^{-1}$ for shrublands and $M_{grass}=1.8$ year$^{-1}$ for grasslands.}. Little work has been done to systematically assess the sensitivity of the standard sequence prediction to model parameters. Such an assessment is important to understanding under what circumstances the sequence is predicted to be an early-warning sign.

Gowda~\etal~\citep{Gowda:2014cd} demonstrate that, in certain limits of model parameter values, pattern sequences can be studied analytically using bifurcation theory. They compute the coefficients of equations which describe the amplitudes of Fourier modes on a 2D hexagonal lattice near a pattern-forming instability. These equations can provide a reduced description of dynamics near the instability. In limits of the model parameter values where transitions between patterns all occur in a regime of weak nonlinearity, their analysis shows that the coefficient of the quadratic-order term in these equations affects the sequence that occurs. If this coefficient changes its sign from negative to positive as a precipitation parameter is decreased in value, an analog of the standard sequence occurs in certain cases. Otherwise, alternative sequences, such as one consisting only of spot patterns, can occur. Based on a preliminary numerical investigation of the model by von Hardenberg~\etal~\citep{vonHardenberg:2001bka}, Gowda~\etal~\citep{Gowda:2014cd} speculate that this coefficient also encodes information about pattern sequences that occur in more fully nonlinear cases.

Here, we ask whether the standard sequence can be identified in the model by Rietkerk~\etal~\citep{Rietkerk:2002ufa} using the sign of the quadratic coefficient of the hexagonal lattice amplitude equations. Specifically, we investigate whether the standard sequence occurs at parameter values where the quadratic coefficient changes its sign from negative to positive as the precipitation parameter decreases. We calculate the amplitude equation coefficients across a range of parameter values, and evaluate these coefficients at two values of the precipitation parameter which correspond to pattern-forming bifurcation points. We conduct numerical simulations of the model over the same parameter range in order to identify the pattern sequences that occur. We compare the results of the analysis with those of the simulations in order to address whether the standard sequence in this model is signalled by the quadratic coefficient changing signs as the precipitation parameter decreases. Of particular interest is whether this is true in regimes where the weakly nonlinear analysis provides no direct information about pattern stability.

If coefficients from bifurcation analysis can serve as proxies for the standard sequence, then this would allow for more efficient exploration of a model's parameter space than by direct numerical simulation. This is important for assessing the robustness of the standard sequence in 3-field models, such as the models by Rietkerk~\etal~\citep{Rietkerk:2002ufa} and Gilad~\etal~\citep{Gilad:2007eh,Zelnik:2013iv}, which depend on a large number of non-dimensional parameters. In this paper, we find that a natural limit of the Rietkerk~\etal~model parameter values permits analytical insight into the model terms which set the sign of the quadratic coefficient. This limit corresponds to a surface water diffusion rate which is much larger than the rate of surface water infiltration as well as the rate of biomass dispersal. Studying the model in this limit allows us to comment on the implications of our numerical study on an ecologically relevant portion of the model parameter space.

This paper is organised as follows. Section~\ref{sec:background} gives background information about the model by Rietkerk~\etal~and the bifurcation-theoretic analysis used in our study of patterned states. Section~\ref{sec:methods} describes the methods used in our study, specifying the parameter spaces explored and summarising the analytic and numerical methods used to explore these parameter spaces. Section~\ref{sec:results} describes the results of the analytic and numerical parameter study, as well as the results of additional analysis performed in a distinguished parameter limit (described in the appendices). Section~\ref{sec:discussion} discusses the implications of our study.

\section{Background}\label{sec:background}
\subsection{Model by Rietkerk~\etal \ (2002)}\label{sec:model}
We study the PDE vegetation model by Rietkerk~\etal~\citep{Rietkerk:2002ufa} (R02), which consists of three fields: surface water $h$, soil water $w$, and plant biomass $n$. Using the non-dimensional form given by Zelnik~\etal~\citep{Zelnik:2013iv}, the model is written as
\begin{equation}
	\begin{aligned}
		\partial_t h = & \underbrace{p}_{\text{precip.}} - \underbrace{I(n)h}_{\text{infil.}} + \underbrace{D_h \nabla^2 h}_{\text{diffusion}},\\
		\partial_t w = & - \underbrace{\nu w}_{\text{evap.}} + \underbrace{I(n) h}_{\text{infil.}} - \underbrace{\gamma G(w)n}_{\text{transp.}} + \underbrace{D_w \nabla^2 w}_{\text{diffusion}}, \\
		\partial_t n = & - \underbrace{\mu n}_{\text{mort.}} + \underbrace{G(w)n}_{\text{growth}} + \underbrace{\nabla^2 n}_{\text{dispersal}},
	\end{aligned}
	\label{eq:R02}
\end{equation}
where
\begin{align*}
	I(n) &= \alpha \frac{n+f}{n+1} \text{ and } G(w) = \frac{w}{w+1}.
\end{align*}
In this model, precipitation is a constant input to the surface water field. Surface water infiltrates and becomes soil water. The infiltration rate (i.e. the conversion rate of surface water to soil water) increases in the presence of biomass via the function $I(n)$ to model the increased permeability of the soil due to plant roots. $I(n)$ saturates to $\alpha$ as $n \to \infty$. Water leaves the soil via evaporation, and is also transpired by plants. The growth rate of biomass is directly proportional to the transpiration rate, and increases with the availability of soil water via the saturating function $G(w)$. Together, these terms make a positive feedback between infiltration and biomass growth: biomass growth increases with soil water, soil water increases with infiltration, and the infiltration rate increases with biomass.

This model includes surface and soil water diffusion terms. Plant dispersal, which encompasses seed dispersal and clonal growth, is also modelled using a diffusion term. The diffusion terms are in two spatial dimensions (2D), i.e. $\nabla^2 = \partial^2 / \partial x^2 + \partial^2 / \partial y^2$. Surface water diffusion is typically assumed to occur much more rapidly than soil water diffusion, so $D_h \gg D_w$. Among three-field PDE vegetation models, soil water diffusion and plant dispersal have been modelled as occurring on either similar~\citep{Rietkerk:2002ufa} or different~\citep{Gilad:2007eh} length and time scales with $D_w \geq 1$. An advection term present in the original form of R02 is neglected here, because the focus of our investigation is on flat-terrain patterns. The dynamics of water on a slope modelled via advection break the symmetry that causes 2D patterns such as gaps or spots at pattern onset.

\begin{figure}[!t]
	\centering \includegraphics[width=0.52\columnwidth]{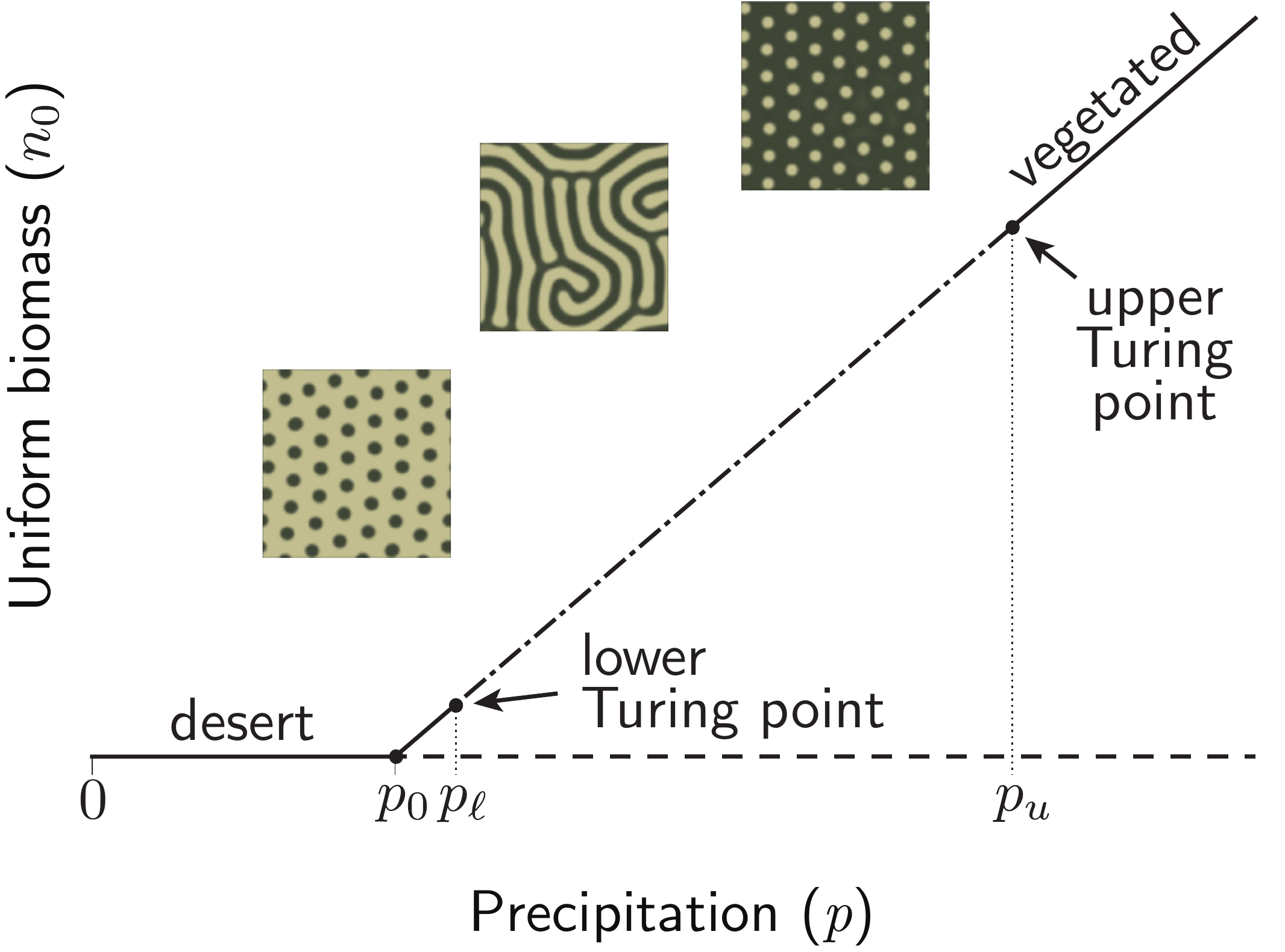} \caption{\small Schematic diagram depicting the uniform steady state solutions of R02~\eqref{eq:R02}, with insets showing examples of patterned states occurring at different values of $p$. The uniform desert state is stable on the interval $p \in [0,p_0)$, and the uniform vegetated state is stable to spatially uniform perturbations for $p > p_0$. The vegetated state is unstable to spatially periodic perturbations at a range of wavelengths on the interval $p \in (p_{\ell}, p_u)$. We refer to the endpoints $p_{\ell}$ and $p_u$ as the lower and upper Turing points, respectively.}\label{fig:R02_bifdiag}
\end{figure}

In general, the form of the growth term varies between models~\citep{Klausmeier:1999woa,Gilad:2007eh}, and it determines the number of uniform steady state solutions that occur for a given system. For R02, the rate of biomass growth depends linearly on the amount of biomass. The growth rate is also a saturating function of soil water, so that it is linear in the amount of soil water for small values of this variable, and constant for large values. This growth rate permits two spatially uniform steady state solutions, which satisfy the equations
\begin{align*}
	0 = &\ p - I(n)h = p - \alpha \frac{n+f}{n+1}h,\\
	0 = &\ - \nu w + I(n) h - \gamma G(w) n = - \nu w + \alpha \frac{n+f}{n+1} h - \gamma \frac{w}{w+1}n, \\
	0 = & \left(-\mu + G(w)\right)n = \left(-\mu + \frac{w}{w+1} \right).
\end{align*}
One solution, $(h,w,n) = (p/f \alpha, p/\nu, 0)$, represents a zero-biomass desert state. The other solution $(h,w,n) = (h_0,w_0,n_0)$ represents a vegetated state with nonzero biomass,
\begin{align}\label{eq:uniform_veg_ss}
	h_0 = \frac{p}{I(n_0)}, \, w_0 = \frac{\mu}{1-\mu}, \, n_0 = \frac{1}{\gamma \mu} \left(p - \frac{\nu \mu}{1-\mu} \right),
\end{align}
for which $n_0 > 0$ when $p > \mu \nu /(1-\mu) \equiv p_0$. We note that the soil water $w_0$ in the vegetated state depends only on the mortality parameter $\mu$, and not on the precipitation level $p$. A diagram of uniform steady state biomass as a function of the precipitation parameter is shown in Figure \ref{fig:R02_bifdiag}. The desert state is stable to spatially uniform perturbations at low values of precipitation ($0 \leq p < p_0$), while the vegetated state is stable to such perturbations at higher values of precipitation ($p > p_0$). These steady states exchange stability in a transcritical bifurcation at $p = p_0$. 

For a range of precipitation values, indicated by the interval $p \in (p_{\ell}, p_u)$, the vegetated state may be unstable to spatially periodic perturbations at a range of wavelengths. When this is the case, pattern-forming instabilities occur at $p_{\ell}$ and $p_u$ via the Turing mechanism~\citep{Turing:1952vn}, and we refer to these points as the lower and upper Turing points respectively. These instabilities result in the formation of spatial patterns, which are eventually stabilised by nonlinearities in the system.

As models of dryland vegetation-water interactions, R02 and related models are conceptual in the sense that they are derived using simple mathematical assumptions for the forms of rate and transport terms. The predictions of the model are not considered robust in regimes where the results of calculations are sensitive to a modest variation in the model parameters or in the exact form of the feedbacks used to model interactions. We explore R02 across a range of parameter values to assess the robustness of model behaviours to parameters.

We consider variations of the infiltration rate parameters, $f$ and $\alpha$. As biomass increases, the infiltration rate approaches $\alpha$. For low values of biomass, infiltration occurs at the lower rate $\alpha f$, with $0 < f < 1$. The parameter $f$ controls the strength of the infiltration feedback, where $f = 0$ corresponds to maximal feedback and $f = 1$ corresponds to no feedback. Some soil types have high infiltration rates and are unaffected by the presence of vegetation, while others have low infiltration rates that are more readily influenced. Because of this, it is appropriate to consider a range of $f$ and $\alpha$ values in a parameter exploration of R02. We also consider variations of the surface water diffusion rate, $D_h$. Turing instabilities in R02 occur when plant dispersal and surface water diffusion occur on different scales. The ratio between surface diffusion and plant dispersal rates, $D_h$, is central to pattern formation in this model, and thus it is varied over a wide range in the investigations in this paper.

\subsection{Analysis of patterns and transitions}~\label{sec:bkgd:analysis}
Pattern formation in R02 and related PDE models is a nonlinear phenomenon and can be studied analytically via bifurcation theory only in certain limits~\citep{Hoyle:2006ur,Cross:2009vp}. One such limit occurs in the vicinity of a Turing point~\citep{Turing:1952vn,Cross:2009vp}. At a Turing point, a uniform steady state has a zero linear growth rate (is neutrally stable) for Fourier mode perturbations of a particular wavelength. We refer to this wavelength as the critical wavelength $\lambda_c$, corresponding to a critical wave number $q_c = 2 \pi/\lambda_c$. The uniform vegetated state of the R02 model can be unstable to Fourier mode perturbations over an interval delimited by two Turing points, which we refer to as the lower and upper Turing points, $p_{\ell}$ and $p_u$ respectively. In this investigation, we analyse patterned states via bifurcation theory near these two points by considering only Fourier modes of a solution on a 2D hexagonal lattice~\citep{Hoyle:2006ur,Cross:2009vp}. The lattice is constructed so that Fourier modes associated with the critical wavelength grow in a neighbourhood of the Turing point, while all others are linearly damped~\citep{Judd:2000wma}. A system of six ordinary differential equations (ODEs) for the Fourier mode amplitudes and their complex conjugates can be written to describe the local pattern-forming dynamics of these critical modes. We study the steady states of these ODEs to gain insight into a system's pattern-forming behaviour near the Turing points.

The critical Fourier modes on a hexagonal lattice take the form
\begin{align}\label{eq:crit_f_modes}
	z_1(t) e^{i\mathbf{q_1}\cdot\mathbf{x}}+z_2(t) e^{i\mathbf{q_2}\cdot\mathbf{x}}+z_3(t) e^{i\mathbf{q_3}\cdot\mathbf{x}}+c.c.,
\end{align}
where $z_1, z_2,$ and $z_3$ are time-dependent complex amplitudes and $c.c.$ denotes the complex conjugates of these modes. The vectors $\mathbf{q_1}$, $\mathbf{q_2}$, and $\mathbf{q_3}$ are wave vectors such that
\begin{equation*}
	\mathbf{q_1} = q_c(1,0), \ \mathbf{q_2} = q_c(-1/2,\sqrt{3}/2), \ \mathbf{q_3} = -(\mathbf{q_1}+\mathbf{q_2}),
\end{equation*}
where $q_c$ is the critical wave number. By assuming that pattern dynamics near a Turing point are well-approximated by the modes~\eqref{eq:crit_f_modes} as a small perturbation to the local uniform steady state, truncated equations for the amplitude dynamics are found to be~\citep{Hoyle:2006ur, Golubitsky:2012ud}
\begin{equation}
	\begin{aligned}
		\dot{z}_1 = m z_1 &+ a \bar{z}_2 \bar{z}_3 - \left(b \vert z_1 \vert^2 + c (\vert z_2 \vert^2 + \vert z_3 \vert^2)\right)z_1,\\
		\dot{z}_2 = m z_2 &+ a \bar{z}_1 \bar{z}_3 - \left(b \vert z_2 \vert^2 + c (\vert z_1 \vert^2 + \vert z_3 \vert^2)\right)z_2,\\
		\dot{z}_3 = m z_3 &+ a \bar{z}_1 \bar{z}_2 - \left(b \vert z_3 \vert^2 + c (\vert z_1 \vert^2 + \vert z_2 \vert^2)\right)z_3. \label{eq:ampeqns}
	\end{aligned}
\end{equation}
These equations are considered to be valid in a regime of weak nonlinearity in a neighbourhood of the Turing point. In this regime, the amplitudes are small so that all three terms can contribute to dynamics on the same order. The linear coefficient $m$ is equivalent to the growth rate of the critical modes obtained by linearising the original system, so that it is small near the Turing point and exactly zero at that point. The quadratic coefficient $a$ must be small for any steady state solutions to~\eqref{eq:ampeqns} to be stable~\citep{Golubitsky:2012ud}, and must be on the order of the size of the amplitude for the quadratic term to balance with the linear and cubic terms. The cubic coefficients $b$ and $c$ are saturating terms, and also affect the stability of steady state solutions of~\eqref{eq:ampeqns}.

Hexagon and stripe steady state solutions of~\eqref{eq:ampeqns} are summarised in Table~\ref{tab:amp_solns_eigs}. Hexagon steady states correspond to equal, real-valued amplitudes, with gaps being negative and spots positive. Stripes correspond to the case of only one nonzero amplitude. The stability of these small-amplitude solutions to perturbations on the hexagonal lattice is dictated by the eigenvalues listed in Table~\ref{tab:amp_solns_eigs}. Of primary importance to this study is the result that $a < 0$ is a necessary condition for the stability of gaps, and $a > 0$ for spots. The quantities $b$, $c-b$ and $b + 2c$ are also related to stability and we refer to these in later interpretation of results.

\begin{table}[!t]
	\caption{Equations for steady state solutions to~\eqref{eq:ampeqns}, and distinct eigenvalues for these solutions. $\mathbf{z} = (z_1,z_2,z_3)$, and $x_s, x_h >0$.}\label{tab:amp_solns_eigs}
	\begin{tabularx}{\columnwidth}{r X l}
		\hline  & Branching equation & Distinct eigenvalues \\
		\hline Gaps ($H^-$): & $\mathbf{z} = -(x_h, x_h, x_h)$ & $3a x_h$,  $2a x_h + 2(c-b)x_h^2$, \\
		 & $0=m - a x_h - (b+2c)x_h^2$   & $-a x_h - 2(b+2c)x_h^2$, 0 \\ \\
		Stripes ($S$): & $\mathbf{z} = (x_s, 0, 0)$  & $-2bx_s^2$, $-(c-b)x_s^2-a x_s$, \\
		& $0 =m - b x_s^2$ &  $-(c-b)x_s^2+a x_s$, 0 \\\\
		Spots ($H^+$): & $\mathbf{z} = (x_h, x_h, x_h)$ & $-3a x_h$, $-2a x_h + 2(c-b)x_h^2$, \\
		& $0=m + a x_h - (b+2c)x_h^2$ & $a x_h - 2(b+2c)x_h^2$, 0 \\
		\hline
	\end{tabularx}
\end{table}

\begin{figure}
	\centering \includegraphics[width=0.52\columnwidth]{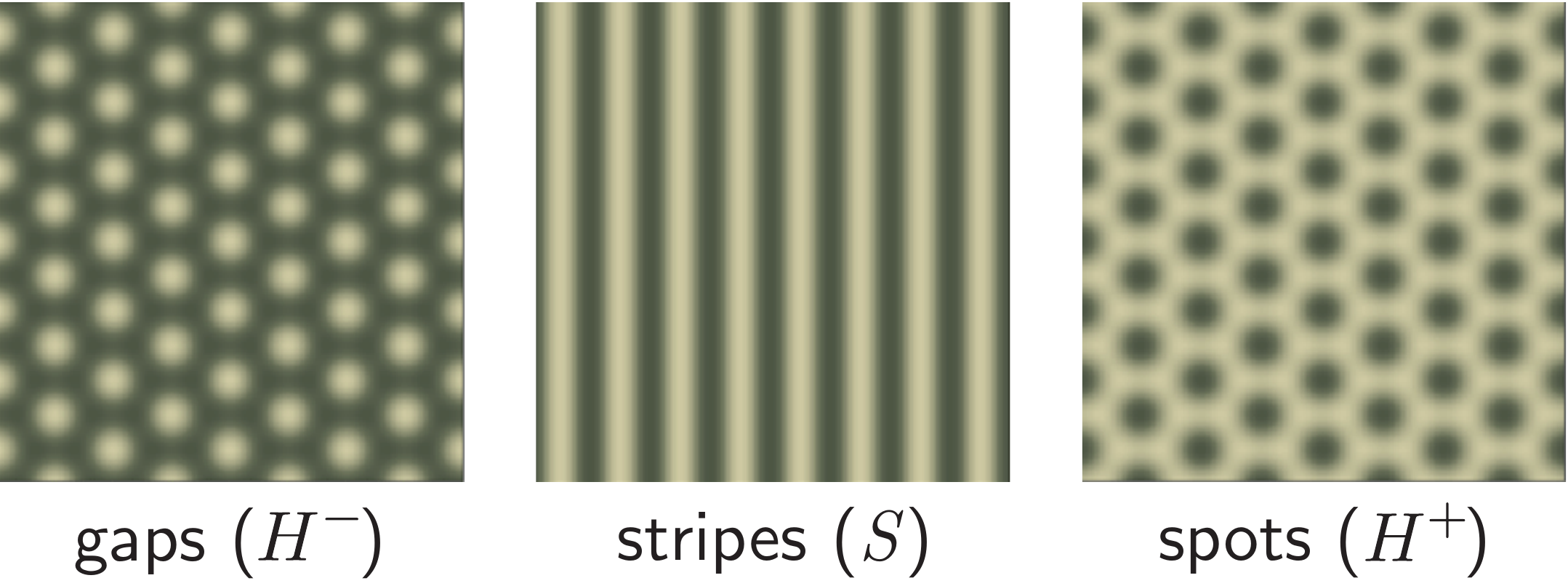} \caption{\small Examples of hexagon ($H^-$ and $H^+$) and stripe ($S$) patterns on a 2D hexagonal lattice. We idealise gaps as $H^-$ patterns, labyrinths as $S$, and spots as $H^+$. } \label{fig:hex_lattice_patterns}
\end{figure}

Gap, labyrinth, and spot-patterned states have previously been observed in R02~\citep{Rietkerk:2002ufa}. Examples of these states are shown in Figure~\ref{fig:R02_bifdiag}. We idealise these states as hexagon and stripe steady states on the hexagonal lattice as in~\citep{LeJeune:2004bma, Gowda:2014cd} (Figure \ref{fig:hex_lattice_patterns}). In general, the lower and upper Turing points occur sufficiently far from one another that it is not possible to study transitions between these states via local bifurcation theory in scenarios of changing precipitation. However, Gowda~\etal~\citep{Gowda:2014cd} consider the special limit in which two Turing points occur near to one another (near to a degenerate Turing point), resulting in pattern transitions that can be analysed via bifurcation theory. In a generic analysis, they find that an analog of the standard ``gaps $\to$ labyrinth $\to$ spots'' sequence is one of many possible sequences that can occur as precipitation decreases. They also find that the value of the quadratic coefficient $a$ at the lower and upper Turing points may serve as a proxy for the apparent pattern sequence. Given appropriate conditions on the cubic coefficients $b$ and $c$ that allow the stability of hexagons and stripe solutions to~\eqref{eq:ampeqns}, they find that if $a < 0$ at the upper Turing point, gaps appear at the start of the sequence. If $a > 0$ at the lower Turing point, then spots appear at the end of the sequence. If these two conditions occur together, then some analog of the standard sequence can occur.

Gowda~\etal~\citep{Gowda:2014cd} speculated that the quadratic coefficient $a$ can be used to obtain information about pattern sequences outside the limited setting of a degenerate bifurcation problem. They conducted a bifurcation analysis and small set of numerical simulations on the model by von Hardenberg~\etal~\citep{vonHardenberg:2001bka} (VH01). VH01 is a two-field PDE vegetation model with cross-diffusion which features two Turing bifurcations on a uniform vegetated steady state (similar to R02). Varying a diffusion coefficient and leaving all other parameters of VH01 fixed, Gowda~\etal~\citep{Gowda:2014cd} found that $a < 0$ at the upper Turing point and $a > 0$ at the lower Turing point in a regime where the diffusion coefficient is sufficiently large. Where stable solutions to the amplitude equations exist within this regime, the signs of the $a$ coefficient at the Turing points allow pattern sequences to begin with stable gaps and to end with stable spots as precipitation decreases. In between these patterns, small-amplitude stripe patterns appear in both analysis and numerical simulations. Where the cubic coefficients $b$ and $c$ prevent the stability of small-amplitude patterns, disordered gaps, labyrinths and spots states were again observed with decreasing precipitation in a numerical simulation. This observation led Gowda~\etal~\citep{Gowda:2014cd} to speculate that the standard sequence occurs in regimes of the model parameter space demarcated by the $a$ coefficient taking a negative sign at the upper Turing point and a positive sign at the lower Turing point, regardless of whether the amplitude equations~\eqref{eq:ampeqns} predict stable patterned states in their small-amplitude regime of validity. Here, we examine whether this holds in the R02 model.

\section{Methods}\label{sec:methods}
We ask whether the quadratic coefficient $a$ of the amplitude equations~\eqref{eq:ampeqns} evaluated at the upper and lower Turing points signals the appearance of the standard sequence in the model by Rietkerk~\etal~\citep{Rietkerk:2002ufa} (R02). To answer this, we conducted bifurcation analyses and numerical simulations across two relevant parameter spaces of the model. To investigate which model terms and parameters influence the sign of the quadratic coefficient, we conducted additional analysis in an ecologically relevant distinguished limit of the model parameters.

\subsection{Model parameters}\label{sec:methods_params}
We studied variations of the non-dimensional R02 parameters given by Zelnik~\etal~\citep{Zelnik:2013iv}, which are based on the dimensioned parameter values estimated by Rietkerk~\etal~\citep{Rietkerk:2002ufa}. These parameter values are summarised in Table~\ref{tab:params}. The $f$ parameter in R02 controls the strength of the infiltration feedback, and is bounded between $0$ and $1$. The default value given in~\citep{Zelnik:2013iv} is $f = 0.2$. We used $f \in [0.1, 0.9]$ in numerical simulations. The $\alpha$ parameter controls the rate of infiltration, and can plausibly take on a large range of values depending on the soil type that is modelled. The default value given in~\citep{Zelnik:2013iv} is $\alpha = 0.4$, and we considered $\alpha \in [10^{-1}, 10^{3}]$. The $D_h$ parameter is the ratio of the surface water diffusion rate and the biomass dispersal rate. Rietkerk~\etal~\citep{Rietkerk:2002ufa} use $D_h = 10^3$ for R02 and Zelnik~\etal~\citep{Zelnik:2013iv} use $D_h = 10^4$ for the corresponding parameter value in a simplified version of the model by Gilad~\etal~\citep{Gilad:2007eh}. We varied $D_h \in [10^{0.5}, 10^{4}]$. To consider the dependence of results on co-variation of parameters, we studied the $\alpha$-$f$ and $D_h$-$f$ parameter spaces. In additional analysis described in the appendices and summarised in Section~\ref{sec:results}\ref{sec:results_quad_analysis}, we considered an ecologically relevant limit where $D_h \gg \alpha$. Our parameter exploration was conducted primarily in this limit, since $\alpha$ was held fixed at $0.4$ while $D_h$ was varied, and $D_h$ was held fixed at $10^3$ while $\alpha$ was varied.

\begin{table*}
	\caption{Parameters given by Zelnik~\etal~\citep{Zelnik:2013iv} for the R02 model, and parameters varied in this study.}\label{tab:params}
	\begin{tabularx}{\textwidth}{cXccc}
		\hline  & Interpretation & Value in~\citep{Zelnik:2013iv} & Constraints & Variation studied\\
		\hline $\mu$ & mortality rate & 0.5 & $0<\mu<1$ & ---\\
		$\alpha$ & infiltration rate & 0.4 & $\alpha > 0$ & $10^{-1}$ -- $10^3$\\
		$f$ & infiltration feedback strength & 0.2 & $0<f<1$ & $0.1$ -- $0.9$ \\
		$\nu$ & evaporation rate & 0.4 & $\nu > 0$ & ---\\
		$\gamma$ & transpiration rate & 0.1 & $\gamma > 0$ & ---\\
		$D_w$ & ratio of soil water diffusion rate to biomass diffusion rate & 1 & $1 \leq D_w < D_h$ & ---\\
		$D_h$ & ratio of surface water diffusion rate to biomass diffusion rate & $10^3$ & $D_h > D_w$ & $10^{0.5}$ -- $10^4$ \\
		\hline
	\end{tabularx}
\end{table*}

\subsection{Amplitude equation calculations}
We computed coefficients of the amplitude equations~\eqref{eq:ampeqns} for R02 using the procedure outlined in Judd~\etal~\citep{Judd:2000wma}, which takes a perturbative approach to obtaining these coefficients for a 2-field reaction diffusion system. These coefficients are written as expressions of the reaction functions and diffusion parameters. We adapted the Judd~\etal~\citep{Judd:2000wma} procedure for a three-field reaction-diffusion system to obtain expressions for the amplitude equation coefficients; the aspect of this procedure specific to the calculation of the quadratic coefficient $a$ is described in Appendix~\ref{apdx:quadcoeff_calc}. The values of the amplitude equation coefficients $a$, $b$, and $c$ are computed at the lower and upper Turing points via~\textit{Mathematica}. These calculations are performed on a grid of points in the $\alpha$-$f$ and $D_h$-$f$ parameter spaces. The results of these calculations were verified at a few non-degenerate points using a centre manifold reduction approach (the general approach is described in~\citep{Wiggins:2003wd}). A Nelder-Mead minimisation library in \textit{Mathematica} was used to find roots of the quantities $a$, $b$, $c-b$, and $b+2c$, which are relevant to assessing the stability of solutions to~\eqref{eq:ampeqns}.

We also calculated the scaling behaviour of the quadratic coefficient $a$ with respect to the quantity $\delta \equiv I(n_0)/D_h$, where $I(n) = \alpha (n+f)/(n+1)$, and $n_0$ is the uniform vegetated biomass given by~\eqref{eq:uniform_veg_ss}. These calculations are presented in the appendices. We found that the critical wave number and onset parameter value scale in distinct ways at each Turing point with respect to $\delta$ when $\delta$ is sufficiently small. We used these scaling relationships to calculate a closed-form expression for the leading order value of $a$ at each Turing point.

\subsection{Numerical simulations}\label{sec:methods_numerics}
\begin{figure}[!t]
	\centering \includegraphics[width=0.6\columnwidth]{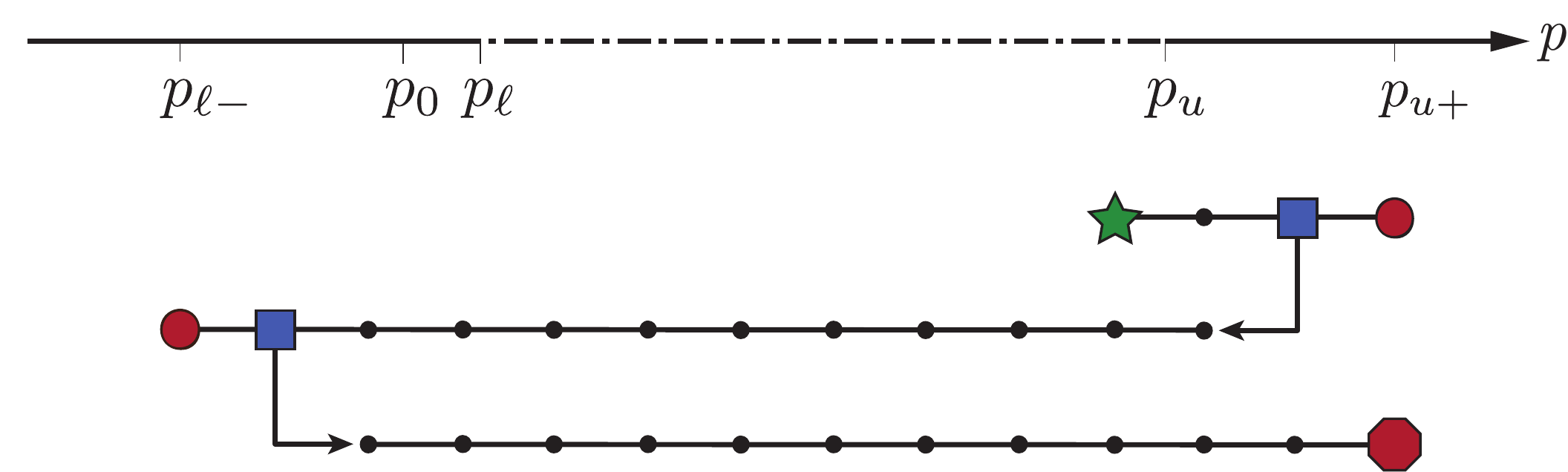} \caption{\small Diagram of numerical simulation procedure. Numerical simulations are run at discrete values of $p$ marked by dots. The procedure is initialised with $p$ just below the upper Turing point $p_u$ (green star) and run forward in time until a steady state is reached. Then $p$ is stepped upward by a small increment and the simulation is once again allowed to reach steady state. This is repeated until patterns lose stability to a uniform vegetated state at $p = p_{u+}$ (red point, right). Using the previous patterned steady state (blue square, right) as an initial condition, $p$ is then stepped downward in the same way until patterns lose stability to a uniform state at $p = p_{\ell-}$ (red point, left). Then $p$ is incremented upward a final time, and the procedure terminates when patterns once again lose stability (red octagon). Note that $p_{u+}$ and $p_{\ell-}$ do not necessarily coincide with the Turing points $p_u$ and $p_{\ell}$, because patterns may persist outside of the Turing instability interval.}\label{fig:numerical_proc}
\end{figure}

We conducted numerical simulations to identify pattern sequences that occur as precipitation decreases in the R02 model. We employed a numerical procedure which simulates an ecosystem undergoing a slow monotonic change in precipitation over time. These simulations were run using grids of parameter values covering regions of the $\alpha$-$f$ and $D_h$-$f$ parameter spaces. The procedure is outlined schematically in Figure \ref{fig:numerical_proc}, and is described in detail in the electronic supplementary material. For each set of parameter values, precipitation is incremented in small discrete steps, and the solution is allowed to reach a steady state between these steps in precipitation. The final state at the previous precipitation value is used as the initial condition for the new precipitation value. The precipitation increment step size was chosen based on the distance between the upper and lower Turing points, $p_u - p_{\ell}$, so that approximately 30-100 end states were saved per simulation. Discrete steps were chosen instead of continuously varying precipitation to avoid transient effects, i.e. simulation results that are sensitive to the rate at which precipitation changes. Simulations were run using the exponential time differencing Runge-Kutta 4 (ETDRK4) scheme~\citep{Cox:2002cga,Kassam:2005jva} modified for 2D systems~\citep{Kassam:2003ws}.

The procedure is constructed to run simulations over the interval of $p$ where patterns are stable and to identify any possible history-dependence (hysteresis) in the pattern sequences. This is accomplished by first incrementing $p$ upward until patterns die out to yield a uniform state. We denote this point of pattern die-off as $p = p_{u+}$. Precipitation is then decremented, simulating a scenario in which an ecosystem slowly becomes more arid. This continues until patterns die out again, which yields another uniform state, which we denote $p = p_{\ell-}$. Precipitation is incremented upward a final time to assess hysteresis in the pattern sequence (i.e. whether the sequence occurs differently when $p$ is slowly increasing versus decreasing). The procedure terminates when patterns die out once more. An approximate interval for the stability of patterns is given by $p \in (p_{\ell-}, p_{u+})$, which contains the Turing instability interval $p \in (p_{\ell}, p_u)$. The Turing instability interval is determined via a linear stability calculation and does not capture the nonlinear stabilisation of patterns. In cases where the amplitude equations~\eqref{eq:ampeqns} predict stable hexagons solutions near a Turing point, these solutions bifurcate in such a way as to be stable outside the Turing interval. When amplitude equation solutions branch away from the Turing interval (e.g. when stripes bifurcate subcritically), these solutions may also stabilise at large amplitude outside the Turing interval.

\section{Results}\label{sec:results}
\subsection{Amplitude equation calculations}\label{sec:results_ampeqns}
\begin{figure}
	\centering  \includegraphics[width=1\textwidth]{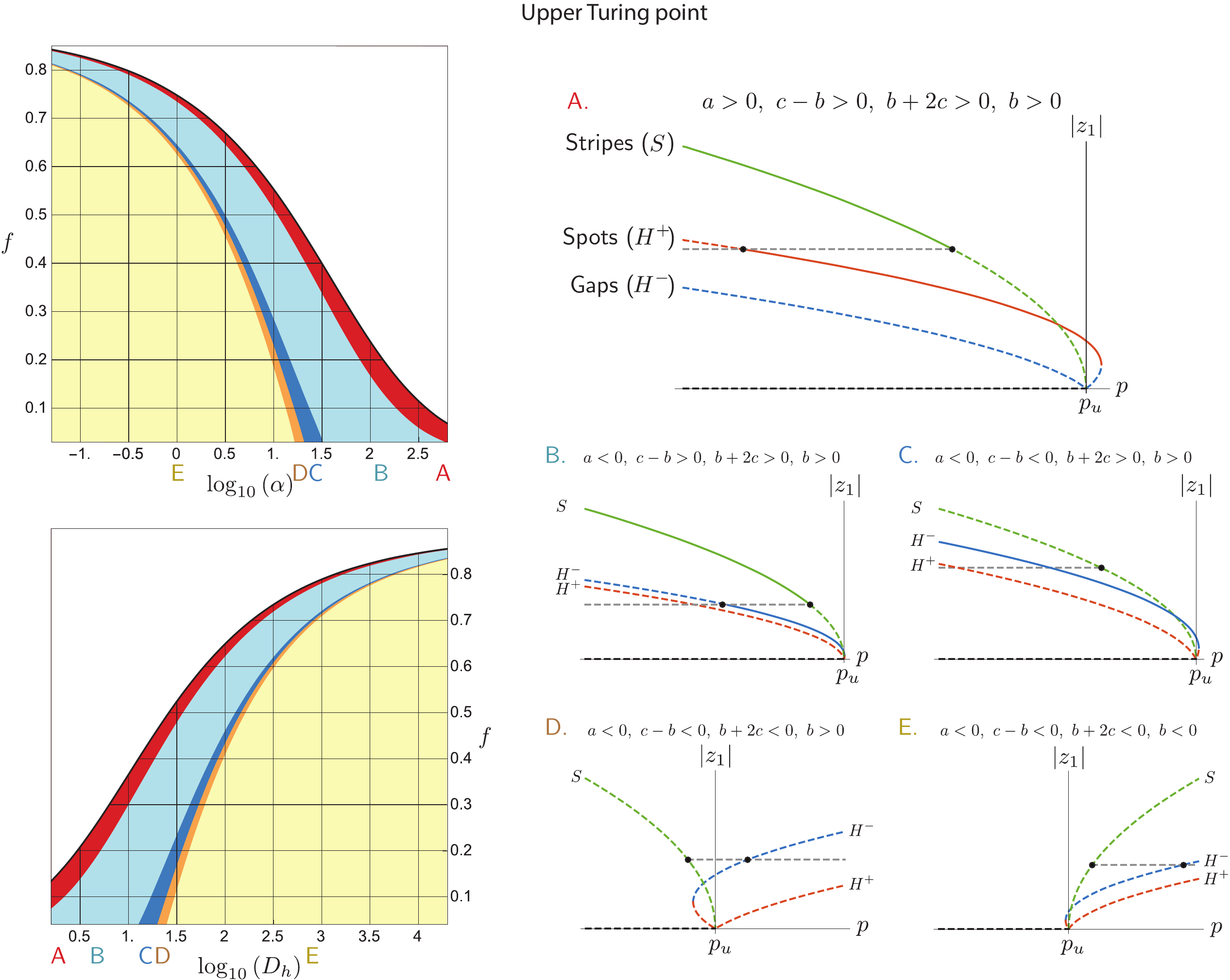} \caption{\small Summary of upper Turing point amplitude equation calculations over $\alpha$-$f$ and $D_h$-$f$ parameter spaces, along with schematic bifurcation diagrams. The coefficients of the amplitude equations~\eqref{eq:ampeqns} are computed, and the curves $a = 0$, $b-c = 0$, $c + 2b = 0$, and $b = 0$ separate the parameter spaces into regions labelled A-E. Qualitatively similar bifurcation structures occur within each region. In the white region, no Turing points occur on the uniform vegetated steady state of R02 and no calculations are performed.} \label{fig:weakly_nonlin_results_upper}
\end{figure}

\begin{figure}
	\centering \includegraphics[width=0.81\columnwidth]{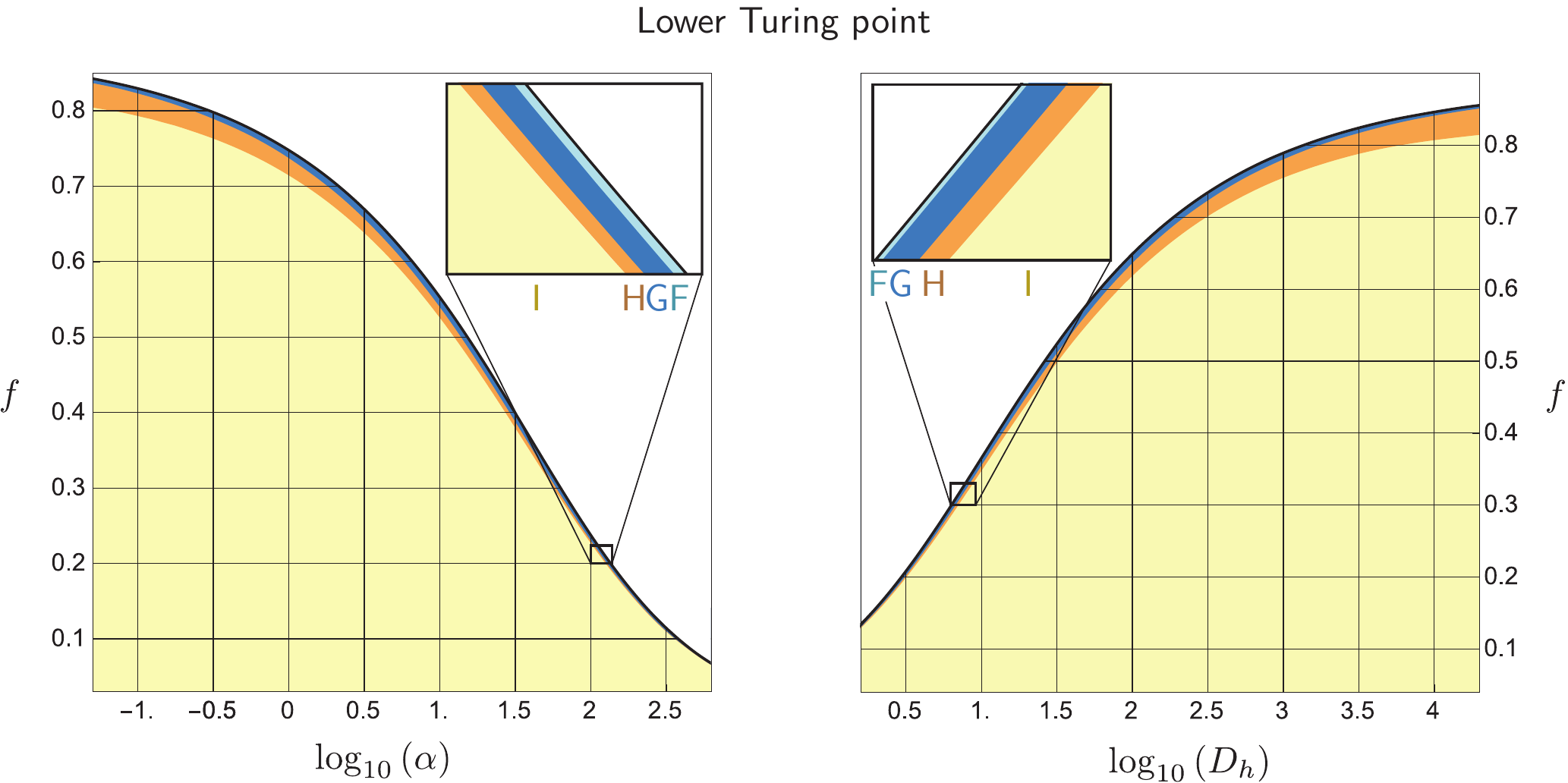} \caption{\small Summary of lower Turing point amplitude equation calculations over $\alpha$-$f$ and $D_h$-$f$ parameter spaces. The coefficients of the amplitude equations~\eqref{eq:ampeqns} are computed, and the curves $c-b = 0$, $b + 2c = 0$, and $b = 0$ separate the parameter space into regions labelled F-I, each of which exhibits a qualitatively distinct bifurcation structure. Bifurcation diagrams applicable to regions F-I resemble diagrams B-E respectively in Figure~\ref{fig:weakly_nonlin_results_upper}, with the roles of gaps and spots exchanged and the solutions reflected so that supercritical branches bifurcate in the direction of increasing precipitation.} \label{fig:weakly_nonlin_results_lower}
\end{figure}

We find that the $\alpha$-$f$ and $D_h$-$f$ parameter spaces of the model by Rietkerk~\etal~\citep{Rietkerk:2002ufa} (R02) can be divided into regions where the amplitude equations~\eqref{eq:ampeqns} give different qualitative predictions. The results of the calculations at the upper Turing point are summarised in Figure~\ref{fig:weakly_nonlin_results_upper}, and the lower Turing point calculations are summarised in Figure~\ref{fig:weakly_nonlin_results_lower}. In the unlabelled white regions, no Turing points occur because the uniform vegetated steady state is stable to spatially periodic perturbations, and no calculations are performed. The black curves separating the white and shaded regions denote a degeneracy of the Turing points, where the upper and lower Turing points come together at a single precipitation value. In the shaded regions below this curve, two Turing points occur on the vegetated state, and analysis is performed at each point.

Each shaded region in Figures~\ref{fig:weakly_nonlin_results_upper} and~\ref{fig:weakly_nonlin_results_lower} is associated with a qualitatively distinct bifurcation diagram applicable to a neighbourhood of the Turing point. The qualitative aspects (e.g. stability, branching direction) of the bifurcation diagrams are determined by the signs of the quantities $a$, $b$, $c-b$, and $b+2c$. These quantities arise from the amplitude equation steady state eigenvalues and branching equations listed in Table~\ref{tab:amp_solns_eigs}. Notably, the sign of the $a$ coefficient serves as a necessary condition for the stability of either small-amplitude gaps or spots solutions. A necessary condition for the stability of gaps is given by $a < 0$, and a necessary condition for the stability of spots is given by $a > 0$. 

For the analyses summarised in Figures~\ref{fig:weakly_nonlin_results_upper} and~\ref{fig:weakly_nonlin_results_lower}, the regions are arrayed similarly in both $\alpha$-$f$ and $D_h$-$f$ parameter spaces. For example, regions A-E in Figure~\ref{fig:weakly_nonlin_results_upper} corresponding to the upper Turing point occur in the same order when varying parameters away from the Turing degeneracy curve (e.g. when increasing $D_h$ compared to decreasing $\alpha$, with $f$ fixed). We find that this occurs because the Turing point calculation and the coefficients $a$, $b$, and $c$ depend on the quantity $\alpha/D_h$, and not $\alpha$ and $D_h$ independently. In the appendices, we show how the Turing point calculation and the quadratic coefficient $a$ depend on $\alpha/D_h$ via the quantity $\delta \propto \alpha/D_h$. Though the amplitude equation coefficients do not depend on $\alpha$ and $D_h$ independently, this does not translate to the invariance of full solutions to R02 for fixed ratios of $\alpha/D_h$. This can be seen in weakly nonlinear solutions, which have linear eigenfunctions that depend on $\alpha$ and $D_h$ independently.

We first interpret the results of bifurcation analysis at the upper Turing point, which are summarised in Figure~\ref{fig:weakly_nonlin_results_upper}. We consider a scenario in which precipitation decreases slowly over time, so that the upper Turing point threshold is crossed from above. The sequence of pattern morphologies observed in such a scenario begins with patterns born near the upper Turing point. The regions in Figure~\ref{fig:weakly_nonlin_results_upper} specify whether the amplitude equations predict a stable patterned state in some neighbourhood of the upper Turing point, and also the morphology of that state. A gap ($H^-$) patterned state stable near the upper Turing point accords with the standard ``$\text{gaps} \to \text{labyrinth} \to \text{spots}$'' sequence prediction.

In region A of Figure~\ref{fig:weakly_nonlin_results_upper}, the quantities $a, \,c-b, \,b + 2c,$ and $b$ are all positive, which allows a stable spot solution ($H^+$) to the amplitude equations in a neighbourhood of the upper Turing point. This analysis predicts that pattern sequences begin with spot patterns in region A of the parameter space, which is inconsistent with the standard sequence. The stripes solution ($S$) can also be stable in region A. It stabilises away from the Turing point, so that spots may transition to stripes as precipitation decreases. However, since the predictions of the amplitude equations break down outside a small neighbourhood of the Turing point, it is uncertain whether this stable stripes solution will manifest in the full system as a successor to spot patterns as precipitation decreases. We expect that as we approach the $a = 0$ boundary of region A, the interval of stability for the spots branch will diminish in size, allowing stable stripes to appear as precipitation decreases.

In regions B and C of Figure~\ref{fig:weakly_nonlin_results_upper}, $a$ is negative and $b + 2c$ and $b$ are positive, which allows a stable gaps solution to the amplitude equations in a neighbourhood the upper Turing point. This predicts pattern sequences that begin with gap patterns, which is consistent with the standard sequence. The stripes solution to the amplitude equations can also be stable in region B, as in region A, since $c-b > 0$. Here, gap patterns may transition to stripes as precipitation decreases. In region C, $c - b< 0$ prevents the stability of stripe steady states. The analysis therefore provides no information in this region about the patterns that may follow gaps as precipitation decreases.

In regions D and E of Figure~\ref{fig:weakly_nonlin_results_upper}, $a, \, c-b,$ and $b+2c$ are negative, which means small-amplitude steady state solutions cannot be stable near the upper Turing point. Regions D and E differ only by the branching direction of the always-unstable stripes solution. The stripes solution branches towards the Turing instability interval (i.e. stripes bifurcate supercritically) for region D, since $b > 0$. The stripes solution branches away from the Turing instability interval (i.e. stripes bifurcate subcritically) for region E, since $b < 0$. In both D and E, the gaps solution branches away from the Turing instability region. Since there are no small-amplitude steady state solutions stable in regions D and E, we cannot directly infer from this analysis what patterned states occur near the Turing point. Here, patterned states of the full system likely arise from more strongly nonlinear behaviour than the states in regions A-C.

Figure~\ref{fig:weakly_nonlin_results_lower} summarises the results of bifurcation analysis at the lower Turing point. Over the entire $\alpha$-$f$ and $D_h$-$f$ parameter spaces, $a$ is positive, which is a necessary condition for the stability of spot solutions to the amplitude equations. Regions F-H all occur in close proximity to the degenerate Turing point curve, while region I fills the majority of the space. Stable solutions to the amplitude equations occur only in regions F and G. In region F, the spots solution to the amplitude equations is stable near the lower Turing point. The stripes solution is also stable away from the Turing point in this region, so that spots may transition to stripes as precipitation increases. In region G, the spots solution is stable near the lower Turing point, but the small-amplitude stripes solution can never be stable. In regions H and I, solutions to the amplitude equations are never stable near the lower Turing point, and differ only in the branching direction of the stripes solution. The stripes solution branches towards the Turing instability region for region H, and away for region I.

At both the upper and lower Turing points, our bifurcation analysis cannot provide direct information about stable patterned states for a large region of the parameter space, where small-amplitude solutions are unstable. The central investigation of this paper is whether the $a$ coefficient, obtained via local analysis, contains information about patterned states near the Turing points in these other regions. In regions B and C of Figure~\ref{fig:weakly_nonlin_results_upper}, it is expected from the analysis that gap patterns are stable near the upper Turing point. We surmise that the same is true in regions D and E, where no solutions to the amplitude equations are stable, but $a < 0$. Similarly, our analysis only shows that spot patterns are stable near the lower Turing point in regions F and G of Figure~\ref{fig:weakly_nonlin_results_lower}. We surmise that spots patterns will be stable near the lower Turing point in regions H and I as well, since $a > 0$ there. We would then expect to see pattern sequences that begin with gaps and end with spots in a scenario of decreasing precipitation over time (i.e. analogs of the standard sequence) in the region of parameter space where $a < 0$ at the upper Turing point and $a > 0$ at the lower Turing point. We find that these conditions are satisfied over nearly all of the studied $\alpha$-$f$ and $D_h$-$f$ parameter spaces, excluding only region A of Figure~\ref{fig:weakly_nonlin_results_upper}. Region A lies adjacent to the Turing degeneracy curve, where the two Turing points approach one another and thus the quadratic coefficients at the Turing points approach the same non-zero value. In this region, $a > 0$ at the both Turing points, and we expect pattern sequences that begin and end with spots.

\subsection{Numerical simulations}
\begin{figure}
	\centering \includegraphics[width=\columnwidth]{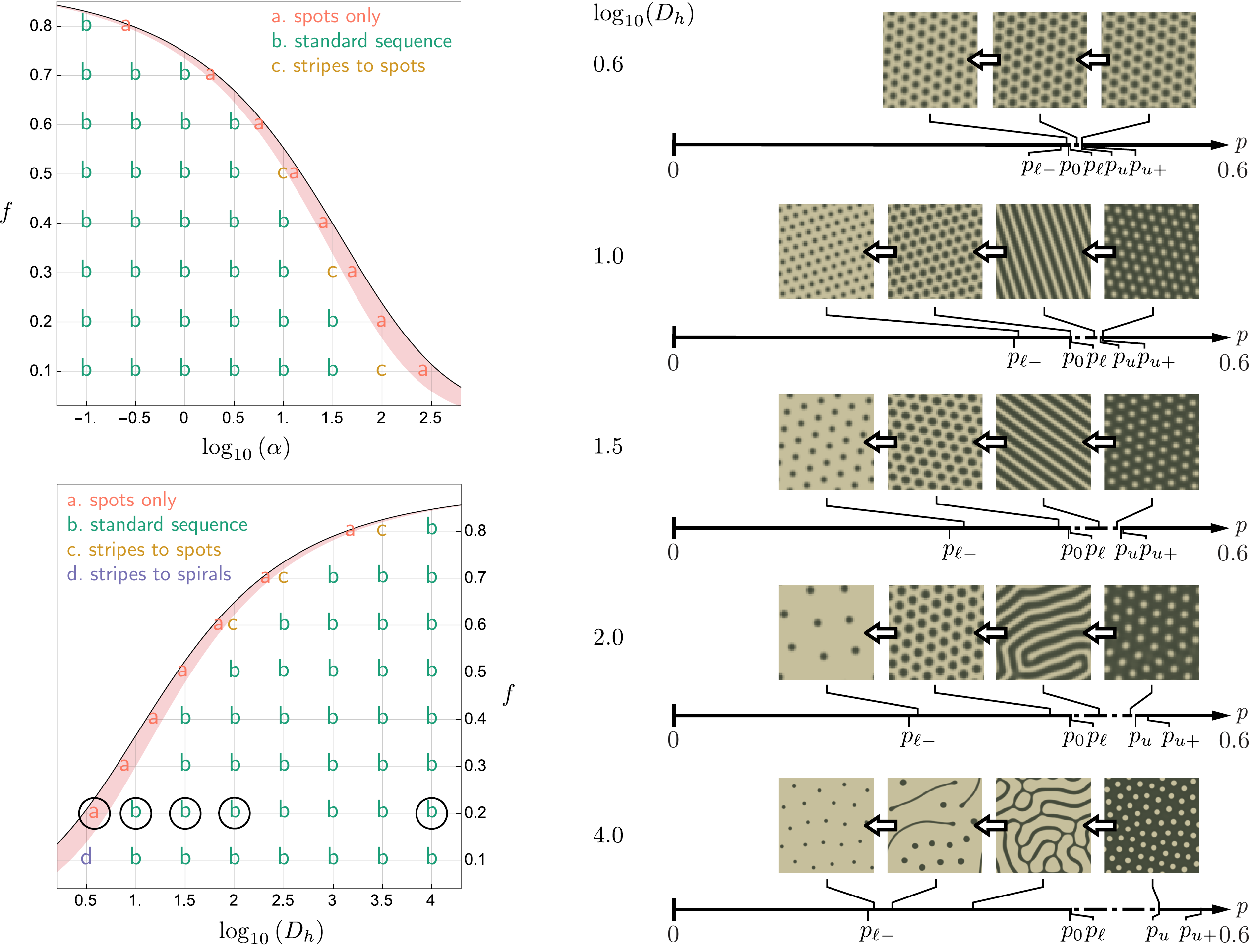} \caption{\small Summary of pattern transitions observed numerical simulations over $\alpha$-$f$ and $D_h$-$f$ parameter spaces in R02, along with representative examples of transitions from numerical simulations at $f =0.2$ and $\log_{10}(D_h) = 0.6$ -- $4.0$. Number lines plot the relative locations of the upper and lower Turing points ($p_u$ and $p_{\ell}$ respectively), the transcritical point ($p_0$), and upper and lower pattern stability boundaries ($p_{u+}$ and $p_{\ell-}$) for the example simulations shown. The parameter values corresponding to the example simulations are circled. Though $p_{\ell}$ and $p_0$ are nearly coincident, the distance between these points is exaggerated to illustrate that $p_{\ell} > p_0$.} \label{fig:main_result}
\end{figure}

In numerical simulations, we find that the quadratic coefficient $a$ of the amplitude equations~\eqref{eq:ampeqns} signals where the standard sequence occurs in the studied parameter spaces of R02. A summary of pattern sequences observed in these numerical simulations is shown in Figure~\ref{fig:main_result}. These simulations were conducted as described in Section~\ref{sec:methods}\ref{sec:methods_numerics} and the electronic supplementary material at sets of parameter values marked by letters, and pattern sequences were identified by visual inspection. For comparison, the region of the parameter space where $a > 0$ at both the upper and lower Turing points (i.e. region A of Figure~\ref{fig:weakly_nonlin_results_upper}) is shaded. Elsewhere, $a < 0$ at the upper Turing point and $a > 0$ at the lower Turing point. From the results of weakly nonlinear analysis described in Section~\ref{sec:results}\ref{sec:results_ampeqns}, we expect pattern sequences beginning with spots (in a scenario of decreasing precipitation) to appear in the shaded region. Outside the thin shaded region, we expect to see analogues of the standard sequence. As expected, only spot patterned states are observed in numerical simulations in the shaded region; in addition, analogues of the standard sequence are primarily observed in simulations elsewhere. 

\begin{figure}
	\centering \includegraphics[width=0.55\columnwidth]{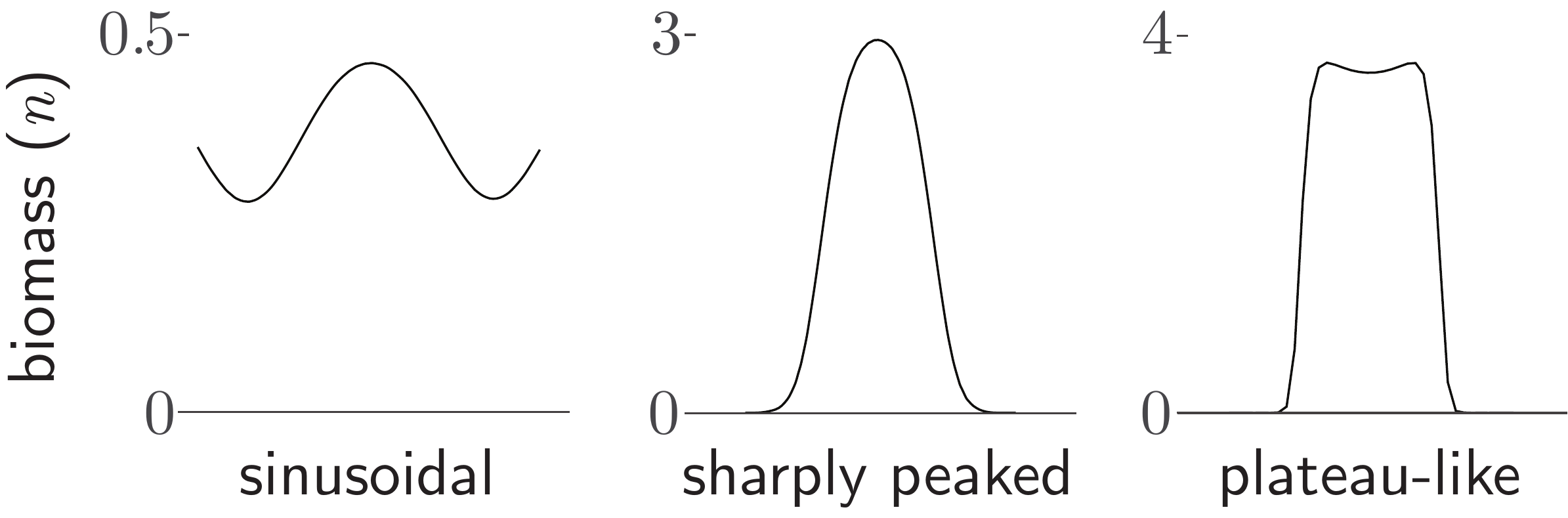} \caption{\small Example profiles of individual spot patterns taken from numerical simulations at $f = 0.2$ and $\log_{10}(D_h) = 0.6$ (sinusoidal), $\log_{10}(D_h) = 2.0$ (sharply peaked), and $\log_{10}(D_h) = 4.0$ (plateau-like). The example sinusoidal profile comes from a spot-patterned state near the upper Turing point. The sharply peaked and plateau-like profiles come from spot-patterned states well below the lower Turing points in their respective simulations.} \label{fig:spot_profiles}
\end{figure}

Examples of the numerical patterned states using $f = 0.2$ and different values of $\log_{10}(D_h)$ are also shown in Figure~\ref{fig:main_result}. The simulation output is accompanied by lines which plot the locations of the upper and lower Turing points, $p_u$ and $p_{\ell}$ respectively, the transcritical point $p_0$, and upper and lower pattern stability boundaries, $p_{u+}$ and $p_{\ell-}$ respectively (defined in Section~\ref{sec:methods}\ref{sec:methods_numerics}). We observed only spot-patterned states in the thin shaded region of Figure~\ref{fig:main_result}. Examples of such states are shown in simulation output from $f = 0.2$ and $\log_{10}(D_h) = 0.6$. Near the upper Turing point, solutions approximately resemble the spots ($H^+$) solution to the amplitude equations~\eqref{eq:ampeqns} shown in Figure~\ref{fig:hex_lattice_patterns}. The profiles of these spot patterns are roughly sinusoidal about the uniform vegetated steady state. An example profile is shown in Figure~\ref{fig:spot_profiles}. At lower values of $p$, spot patterns remain stable. The spacing between spots increases, and the individual spots of vegetation become more sharply peaked, quickly decaying to zero away from the centre of a spot. An example of a sharply peaked profile is also shown in Figure~\ref{fig:spot_profiles}. Patterns other than spots are not observed in simulations conducted in the shaded region. No notable difference in the qualitative appearance of the spot patterns was observed as precipitation increased in discrete steps.

We primarily observed analogues of the standard ``gaps $\to$ labyrinths $\to$ spots'' sequence in the unshaded region of Figure~\ref{fig:main_result}, which agrees with our expectations from analysis. Examples of this sequence in simulation output for $f = 0.2$ and $\log_{10}(D_h)$ ranging from $1.0$ -- $4.0$ are shown in Figure~\ref{fig:main_result}. The sets of simulations at $\log_{10}(D_h) = 1.0$ and $\log_{10}(D_h) = 1.5$ use parameter sets from regions B and C of Figure~\ref{fig:weakly_nonlin_results_upper} respectively, where gaps solutions to the amplitude equations are expected to be stable near the upper Turing point. As precipitation decreases in the simulations, patterns resembling the gaps ($H^-$) solution to the amplitude equations are first observed near the upper Turing point. Gaps then transition to well-ordered stripe patterns in both sets of simulations. As precipitation decreases further, stripes become disordered before transitioning to spot patterns. The sets of simulations at $\log_{10}(D_h) = 2.0$ and $\log_{10}(D_h) = 4.0$ both use parameter sets from region E of Figure~\ref{fig:weakly_nonlin_results_upper}, where no small-amplitude patterns are stable near the upper Turing point. Still, gaps are observed in numerical simulations near the upper Turing point in both sets of simulations. As precipitation decreases, gaps transition directly to disordered labyrinthine stripes. These stripes eventually transition to spots, which take on non-sinusoidal profiles as shown in Figure~\ref{fig:spot_profiles}. Hysteresis occurs in the points of transition between pattern morphologies, and this hysteresis is larger in the simulation with the larger value of $D_h$. The transitions between gaps and labyrinths occur at a lower value of $p$ when decreasing precipitation compared to increasing precipitation. The same applies to the transition between labyrinths and spots. 
 
The simulation examples in Figure~\ref{fig:main_result} demonstrate a trend of increasingly nonlinear behaviour as $D_h$ increases (see also~\citep{Siero:2015ht, Kolokolnikov:2006hi} for similar behaviour in related systems). This trend is generally representative of what we observe in the other simulations when parameters are varied away from the Turing degeneracy curve, e.g. when $\alpha$ decreases. One aspect of the increasing nonlinearity can be accounted for via our bifurcation analysis. The regions in Figure~\ref{fig:weakly_nonlin_results_upper} order the parameter spaces by nonlinearity at the upper Turing point. Moving away from the Turing degeneracy curve, the amplitude equations first predict stable weakly nonlinear patterns in regions A-C, and then imply strongly nonlinear patterns in regions D and E since small-amplitude patterns are unstable. This manifests in simulations as small-amplitude sinusoidal patterns occurring near the upper Turing point when parameters are near to the Turing degeneracy curve, sharply peaked patterns occurring beyond this, and plateau-like patterns occurring when very far away from the curve. Other aspects of increasing nonlinearity are apparent in certain qualitative behaviours observed in the simulations. Patterned states increase in their disorder and begin to exhibit coexistence as distance from the degeneracy curve increases. The interval of pattern stability, $(p_{\ell-}, p_{u+})$, increases in length as well. In particular, $p_{\ell-}$ decreases to extend the length of the interval $(p_{\ell-},p_{\ell})$. As this interval increases with increasing distance from the Turing degeneracy curve, the transition point between stripe and spot states decreases to lower values of precipitation. This causes spot patterns to remain stable at values of precipitation well below the lower Turing point. This implies the stabilisation of a strongly nonlinear patterned state far from the Turing instability interval.

In addition to the standard sequence, we observed a few instances of ``stripes $\to$ spots'' sequences in the unshaded region of Figure~\ref{fig:main_result}. We determined that these are actually instances of the standard sequence, where gaps do not appear in simulations. Our bifurcation analysis indicates that gap solutions to the amplitude equations are stable only very near to the Turing point in these parameter sets. Because our numerical procedure increments precipitation in discrete steps of fixed size, the gaps branch may be bypassed. To test whether gaps can exist stably for parameter sets where ``stripes $\to$ spots'' transitions are observed, we conducted additional numerical simulations, which are described in the electronic supplementary material. In these simulations, gap patterns were assessed to be stable. 

We also observed time-varying spiral wave patterns in one instance of the numerical simulations, at $f = 0.1$ and $\log_{10}(D_h) = 0.5$. We ran additional simulations, described in the electronic supplementary material, at nearby parameter values and found that spirals patterns are confined to values of $D_h$ that are smaller than typically considered ecologically applicable. To our knowledge, there are no previous reports of spiral wave patterns occurring in R02 or any other vegetation model. However, we remark that the waves observed in R02 resemble spiral wave patterns observed in other reaction-diffusion contexts such as chemical reaction systems~\cite{Vanag:2007dt} and models of phytoplankton dynamics~\cite{Malchow:2004ema}.

\subsection{Quadratic coefficient analysis}\label{sec:results_quad_analysis}
In Appendix~\ref{apdx:quadcoeff_calc}, we derive a closed-form expression for the quadratic coefficient $a$. This expression involves derivatives of nonlinear terms in R02, the infiltration term $I(n)h$ and the growth term $G(w)n$, where $I(n) = \alpha (n+f)/(n+1)$ and $G(w) = w/(w+1)$. This expression also involves the null vector $(H_1,W_1,N_1)^T$ and left null vector $(\widetilde{H}_1,\widetilde{W}_1,\widetilde{N}_1)$ of the system linearised about the uniform vegetated steady state $(h_0, w_0, n_0)$~\eqref{eq:uniform_veg_ss}, both of which are defined in Appendix~\ref{apdx:quadcoeff_calc}. Explicitly,
\begin{multline}\label{eq:quad_coeff_full}
	a = \left( I'(n_0) H_1 N_1 + \frac{1}{2} I''(n_0) h_0 N_1^2 \right)\left(\widetilde{W}_1-\widetilde{H}_1\right) \\+ \left(G'(w_0) W_1 N_1 + \frac{1}{2} G''(w_0) n_0 W_1^2 \right)\left(\widetilde{N}_1-\gamma \widetilde{W}_1\right).
\end{multline} 
The infiltration function $I(n)$ is an increasing concave-down function of $n$, and so $I'(n) > 0$ and $I''(n) < 0$. The growth function $G(w)$ is similarly increasing and concave-down with respect to $w$. 

In a natural limit of the parameters which corresponds to $D_h \gg \alpha$, we find in Appendix~\ref{apdx:quadcoeff_analysis} that a negative term dominates quadratic coefficient $a$~\eqref{eq:quad_coeff_full} at the upper Turing point and a positive term dominates at the lower Turing point. These arise through distinct scalings of the critical wavenumber and the onset parameter values with the quantity $\delta \equiv I(n_0)/D_h \propto \alpha/D_h$ at the different Turing points, and are calculated in Appendix~\ref{apdx:wavenumber}. These scalings at onset in turn determine distinct scalings for the left and right null vector components. At the lower Turing point, for $\delta$ sufficiently small, we find that $(H_1,W_1,N_1)^T = (\mathcal{O}(1), \mathcal{O}(1), \mathcal{O}(1))^T$ and $(\widetilde{H}_1,\widetilde{W}_1,\widetilde{N}_1) = (\mathcal{O}(\delta), \mathcal{O}(\delta), \mathcal{O}(1))$. Additionally we find that $n_0 = \mathcal{O}(\delta)$. Then at the lower Turing point, 
\begin{align*}
	a = \frac{2 Q_{\ell}^2 G'(w_0)}{N_{\ell}} N_1^{\ell} + \mathcal{O}(\delta),
\end{align*}
where $Q^2_{\ell}$ and $N_{\ell}$ are the positive scaling constants given by~\eqref{eq:lower_coeff_eqn}. Thus $a$ is positive at leading order at the lower Turing point. This corresponds to spot patterns. Sufficiently far from the degeneracy at which the two Turing points merge, the upper Turing point has the opposite sign. Specifically, at the upper Turing point, we find that $(H_1,W_1,N_1)^T = (\mathcal{O}(\delta^{1/2}), \mathcal{O}(\delta^{1/2}), \mathcal{O}(1))^T$ and $(\widetilde{H}_1,\widetilde{W}_1,\widetilde{N}_1) = (\mathcal{O}(\delta^{1/2}), \mathcal{O}(1), \mathcal{O}(1))$. Then at the upper Turing point
\begin{align*}
	a = \frac{G'(w_0) I''(n_0) h_0 n_0}{\nu + \gamma G'(w_0)n_0} + \mathcal{O}(\delta^{1/2}).
\end{align*}
This expression is negative at leading order since $I''(n)<0$, and thus corresponds to gap patterns. 

In order for this result to hold, we must also be sufficiently far from a Turing degeneracy. The quadratic coefficient at the two Turing points takes on the same sign in a neighbourhood of the degeneracy, marked by the shaded region in Figure~\ref{fig:main_result} (see also the discussion of region A in Section~\ref{sec:results}\ref{sec:results_ampeqns}). The conditions that $D_h \gg \alpha$ and that we are sufficiently far from a Turing degeneracy are not independent. As $f$ increases, the Turing degeneracy occurs at smaller values of $\delta$. This can be observed in Figures~\ref{fig:weakly_nonlin_results_upper}-\ref{fig:main_result}, where $\alpha$ decreasing and $D_h$ increasing correspond to smaller $\delta$. At smaller $\delta$, the regions where $a$ takes the same sign at both Turing points diminish, and less distance from the Turing degeneracy is necessary for the scaling results given above to hold. This can be seen in the narrowing of the shaded regions in Figure~\ref{fig:main_result} with increasing $f$.

\section{Discussion}\label{sec:discussion}
We find that the quadratic coefficient $a$ from a bifurcation analysis divides the studied parameter spaces of the Rietkerk~\etal~\citep{Rietkerk:2002ufa} (R02) model into two regions. In a thin region of the parameter space adjacent to the degenerate Turing point curve, where the Turing points are very close to each other, $a$ is positive at both points. Correspondingly, we observe only spot patterns in numerical simulations. Elsewhere $a$ takes opposite signs at the two Turing points. When this happens, we primarily observe the standard sequence. This strongly suggests that the $a$ coefficient resulting from weakly nonlinear analysis holds predictive value for assessing the nonlinear behaviour of the system. Specifically, it appears to serve as a proxy for the sequence of nonlinear patterns that will manifest for any parameter set of R02.  

Since $a$ is computed analytically, it is possible to trace the influence of model terms and parameter values on the sign of $a$, and thus on the sequence of patterns that are predicted. This presents an approach for comprehensively exploring the full 7-dimensional parameter space of the R02 model. Our analysis of the quadratic coefficient in the appendices shows that in any parameter regime where the surface water diffusion rate $D_h$ is sufficiently large compared to the infiltration rate $\alpha$, the quadratic coefficient takes values consistent with standard sequence.

We believe that $D_h \gg \alpha$ is an appropriate limit for the R02 model. We have treated the parameters $D_h > 1$, $\alpha > 0$ and infiltration feedback strength $0<f<1$ as essentially unconstrained in this study. Our analysis shows, however, that the Turing point calculation is invariant to fixed values of the ratio $\alpha/D_h$ (see Appendix~\ref{apdx:wavenumber}), and that Turing bifurcations only occur for $\alpha/D_h$ sufficiently small. This can be seen in Figures~\ref{fig:weakly_nonlin_results_upper}-\ref{fig:main_result}, where Turing bifurcations only occur for $\alpha$ sufficiently small with $D_h = 10^3$, and for $D_h$ sufficiently large with $\alpha = 0.4$. The amount of separation between $\alpha$ and $D_h$ that is required for Turing points to occur depends on $f$, which can also be seen in Figures~\ref{fig:weakly_nonlin_results_upper}-\ref{fig:main_result}. For values of $f$ greater than 0.6, holding other parameters fixed at default values, $\alpha$ and $D_h$ must be separated by at least two orders of magnitude for Turing points to occur. For any value of $f \in (0,1)$, not much additional separation between $\alpha$ and $D_h$ is required for the quadratic coefficient $a$ to take opposite signs at the Turing points and for the standard sequence to occur.

Moreover, the limit where $D_h \gg \alpha$ is ecologically relevant. For all soil types, we expect $D_h$ to be large, because it represents the ratio of surface water diffusion to biomass dispersal, which occur at quite different scales. Additionally, a global study of the factors associated with the existence of vegetation patterns by Deblauwe~\etal~\citep{Deblauwe:2008if} finds that patterns favour environments with non-sandy soils, where relatively small rates of infiltration allow for substantial water redistribution via surface runoff. For such non-sandy soils, Rietkerk~\etal~\citep{Rietkerk:2002ufa} estimate an infiltration rate and surface diffusion rate that results in $\alpha/D_h \approx 5 \times 10^{-4}$. Holding the other model parameters fixed at their default values, this corresponds to the standard sequence for $99$\% of the range of $f$ over which patterns are present.

Although vegetation patterns tend to occur in non-sandy soils, a notable exception is the fairy circle phenomenon in Namibia. These patterns occur in a sandy soil environment marked by a high infiltration rate~\citep{Zelnik:2015kf}. In empirical comparisons between clayey soils and sandy soils, infiltration rates differ only by a factor of approximately 10-20~\citep{Hillel:2013tn}. In the Namibian system specifically, estimates for the infiltration rate parameter used in the model by Gilad~\etal~\citep{Gilad:2004bp, Gilad:2007eh, Zelnik:2013iv} are at most one order of magnitude larger than those in the non-sandy R02 system\footnote{Getzin~\etal~\citep{Getzin:2014kt} estimate a nondimensional infiltration rate for the model by Gilad~\etal~\citep{Gilad:2004bp, Gilad:2007eh, Zelnik:2013iv} as $\alpha_{Getzin} \equiv A/M = 6,$ where $A$ is a dimensioned infiltration rate and $M$ is a mortality rate. The nondimensional infiltration and mortality rates for R02 are given by Zelnik~\etal~\citep{Zelnik:2013iv} as $\alpha_{R02} \equiv A/(c g_{max}), \ \mu_{R02} \equiv M/(cg_{max})$, where $1/(cg_{max})$ sets the time scale of the nondimensional system. Therefore, $\alpha_{R02} = \alpha_{Getzin} \mu_{R02} < \alpha_{Getzin}$.}~\citep{Getzin:2014kt}. Given similar surface water diffusion rates to R02, this results in the standard sequence over about $95$\% of the range of $f$ for which patterns are present in the R02 model ($\alpha/D_h \approx 5 \times 10^{-3}$).  

We conclude that the standard sequence prediction appears robust to parameter variation in the R02 model. This conclusion is based on evidence that the quadratic coefficient serves as a proxy for the standard sequence, along with our finding that the coefficient takes on values consistent with the standard sequence in the ecologically relevant region of the parameter space. The methodology presented here provides a way forward for assessing the robustness of the standard sequence in other models and formulations. Doing so is an important step towards establishing credibility for the standard sequence as an early-warning sign, since other model formulations may lead to different pattern sequence predictions. The analysis conducted in this paper is specific to the formulation of R02. For instance, R02 is distinctive due to the form of its growth function $G$, which depends only on the soil water field ($G = G(w) = w/(w+1)$). This leads to a soil water value in the uniform vegetated steady state that is constant with respect to precipitation level, and only dependent on the mortality parameter ($w_0 = \mu/(1-\mu)$). This feature simplifies the analysis of this system. Modifications to the growth function through the addition of dependence on biomass (e.g. $G = G(w,n)$), such as in the models by Gilad~\etal~\citep{Gilad:2004bp, Gilad:2007eh, Zelnik:2013iv} and Klausmeier~\cite{Klausmeier:1999woa,vanderStelt:2013wy} alter the structure of the uniform vegetated steady state~\citep{Iams:2015rk}. Because the value of the quadratic coefficient depends on the form of $G$, we may not find the standard sequence to be similarly robust in other systems with different growth functions. Assessing the robustness of the standard sequence to alternative (and possibly more general) model formulations is an important and substantial undertaking, and could be a direction for future studies. 

Not all PDE vegetation models feature two Turing points on the vegetated steady state. For such models our methodology cannot be applied exactly. The Klausmeier Gray-Scott model~\citep{Klausmeier:1999woa,vanderStelt:2013wy} and Simplified Gilad model~\citep{Gilad:2004bp, Gilad:2007eh, Zelnik:2013iv} (in the parameter regime typically studied) have only one Turing point on the vegetated steady state. This corresponds, in some sense, to the upper Turing point in R02. In the analysis of R02 conducted here, the information used to divide the parameter space into regions with different pattern sequences was from the upper Turing point. Because the lower Turing point remained positive throughout the parameter space studied, spot patterns were always observed in simulations at low levels of precipitation. It may be true that in vegetation models with only one Turing point, the sign of the $a$ coefficient at that point is sufficiently informative to predict where the standard sequence occurs in the parameter space. This would represent a remarkable simplification, permitting the efficient analysis of pattern transitions in other PDE vegetation models.
\\\\
{\small
\noindent\textbf{Author contributions} KG, SI, and MS conceived of this study; KG designed and conducted the numerical simulations; KG and YC conducted the amplitude equation calculations; YC and MS conceived of the quadratic coefficient analysis in the appendices; KG, YC, and MS conducted the analysis in the appendices; KG drafted the manuscript with contributions from SI on the model background; YC, SI, and MS gave feedback on the manuscript; All authors gave final approval for publication.\\
\textbf{Competing interests} We have no competing interests.\\
\textbf{Funding} Research was supported in part by NSF DMS-1517416 and by the NSF Math and Climate Research Network (DMS-0940262).\\
\textbf{Acknowledgments} We benefitted from useful conversations at the ``Spatio-Temporal Dynamics in Ecology'' workshop at the Lorentz Centre (Leiden, NL, December 2014), and also from constructive comments by referees of this manuscript.}

\section*{Appendix}
\appendix
\section{Quadratic coefficient calculation}\label{apdx:quadcoeff_calc}
In this appendix, we summarise the calculation of the quadratic coefficient, $a$, of the amplitude equations~\eqref{eq:ampeqns} for the Rietkerk~\etal~\citep{Rietkerk:2002ufa} model (R02). We do so in a manner that illustrates the role played by the nonlinear functions in R02.

We expand R02~\eqref{eq:R02} to quadratic order about the uniform vegetated steady state $(h_0,w_0,n_0)$, which is a function of precipitation $p$ given by~\eqref{eq:uniform_veg_ss}. We take $H = h-h_0, \, W = w-w_0, \, N = n-n_0$:
\begin{align}\label{eq:quad_expansion}
	\frac{\partial}{\partial t} \begin{pmatrix} H\\W\\N \end{pmatrix} = \mathcal{L}\begin{pmatrix} H\\W\\N \end{pmatrix} + \begin{pmatrix}-I_2(H,N) \\ I_2(H,N) - \gamma G_2(W,N) \\ G_2(W,N) \end{pmatrix} + ... .
\end{align}
The linear operator $\mathcal{L}$ is
\begin{align}\label{eq:linearoperator}
	\mathcal{L} =  \begin{pmatrix} -I(n_0) + D_h \nabla^2 & 0 & - I'(n_0)h_0  \\ I(n_0)  & - \nu -\gamma G'(w_0)n_0 + D_w \nabla^2 &  I'(n_0)h_0- \gamma \mu \\ 0 & G'(w_0)n_0  & \nabla^2 \end{pmatrix},
\end{align}
the quadratic order terms $I_2(H,N)$ and $G_2(W,N)$ are
\begin{align}\label{eq:I2andG2}
	I_2(H,N) &= I'(n_0) H N + \frac{1}{2} I''(n_0) h_0 N^2,\\
	G_2(W,N) &= G'(w_0) W N + \frac{1}{2} G''(w_0) n_0 W^2,
\end{align}
and the ellipsis denotes terms of cubic order in $(H,W,N)$. The linear stability of the uniform vegetated state to spatially periodic perturbations with wave number $q$ is determined by substituting $(H,W,N)^{T} = \bm{\xi} e^{i q x} e^{\sigma t}$ into equation~\eqref{eq:quad_expansion} linearised about $H = W = N = 0$. This gives the eigenvalue problem $\sigma \bm{\xi} = J(q^2, p) \bm{\xi}$, where $J(q^2,p)$ is the Jacobian matrix
\begin{align}\label{eq:jacobian}
	J(q^2,p) =  \begin{pmatrix} -I(n_0) - D_h q^2 & 0 & - I'(n_0)h_0  \\ I(n_0)  & - \nu -\gamma G'(w_0)n_0 - D_w q^2 &  I'(n_0)h_0- \gamma \mu \\ 0 & G'(w_0)n_0  & -q^2 \end{pmatrix}.
\end{align}
The Jacobian matrix depends explicitly on the wave number $q$, as well as on the precipitation parameter $p$ through the uniform vegetated steady state $(h_0,w_0,n_0)$. We consider $p$ to be the bifurcation parameter in this analysis. A Turing point occurs at a parameter value $p = p_c$ and wave number $q^2 = q_c^2>0$ for which the maximum real part of an eigenvalue is zero, and all other modes are damped. Necessary conditions for a Turing point $(q_c^2,p_c)$ are given by
\begin{align}
	\text{Det}(J(q_c^2,p_c)) &= 0\label{eq:turingconditions1}, \\ \left.\frac{\partial \text{Det}(J(q^2,p_c))}{\partial q^2}\right\vert_{q = q_c} &= 0.\label{eq:turingconditions2}
\end{align}

We follow a standard procedure~\citep{Judd:2000wma,Hoyle:2006ur,Cross:2009vp} to obtain the quadratic coefficient $a$ of the amplitude equations~\eqref{eq:ampeqns}. We write a small amplitude hexagonal (spots/gaps) solution to~\eqref{eq:quad_expansion} near $(q_c^2,p_c)$ as
\begin{align}\label{eq:small_soln}
	\begin{pmatrix} H \\ W \\ N \end{pmatrix} = \epsilon \left(z(t_1) f(\bm{x}) + c.c. \right) \begin{pmatrix} H_1 \\ W_1 \\ N_1 \end{pmatrix} + \epsilon^2 \bm{U_2} + \mathcal{O}(\epsilon^3),
\end{align}
where $\epsilon \ll 1$, $t_1 = \epsilon t$,
\begin{align*}
	f(\bm{x}) = e^{i\bm{q_1}\cdot\bm{x}} + e^{i\bm{q_2}\cdot\bm{x}} + e^{i\bm{q_3}\cdot\bm{x}},
\end{align*}
$(H_1,W_1,N_1)^T$ is to be determined, and $\bm{U_2}$ is a higher-order term which must be bounded. The wave vectors lie on a 2D hexagonal lattice, with $\mathbf{q_1} = q_c(1,0), \ \mathbf{q_2} = q_c(-1/2,\sqrt{3}/2), \ \mathbf{q_3} = -(\mathbf{q_1}+\mathbf{q_2})$. Plugging~\eqref{eq:small_soln} into~\eqref{eq:quad_expansion} gives the $\mathcal{O}(\epsilon)$ equation
\begin{align*}
	0 = J(q_c^2, p_c) \begin{pmatrix} H_1 \\ W_1 \\ N_1 \end{pmatrix}.
\end{align*}
Since $J(q_c^2,p_c)$ has a zero eigenvalue, we take $(H_1, W_1, N_1)^{T}$ to be the associated right null vector to satisfy this equation. We choose the convention that $H_1^2 + W_1^2 +  N_1^2 = 1$ and $N_1 > 0$.

At $\mathcal{O}(\epsilon^2)$, we have
\begin{align*}
	f(\bm{x}) \frac{\partial z}{\partial t_1} \begin{pmatrix}H_1\\N_1\\W_1\end{pmatrix}  = \mathcal{L} \bm{U_2} + \begin{pmatrix} -I_2(H_1,N_1) \\  I_2(H_1,N_1) - \gamma G_2(W_1,N_1) \\ G_2(W_1,N_1) \end{pmatrix} \left(z(t_1) f(\bm{x}) + c.c. \right)^2.
\end{align*}
The term $\left(z(t_1) f(\bm{x}) + c.c.\right)^2$ generates modes with wave vectors of magnitude $q_c$:
\begin{align*}
	(z(t_1) f(\bm{x}) + c.c.)^2 = 2 \bar{z}^2 \left(e^{i\bm{q_1}\cdot\bm{x}} + e^{i\bm{q_2}\cdot\bm{x}} + e^{i\bm{q_3}\cdot\bm{x}}\right) + c.c. + ...
\end{align*}
These modes result in secular terms in the solution~\eqref{eq:small_soln}. To eliminate these terms, we apply the Fredholm alternative theorem to obtain the solvability condition $\partial z/\partial t_1 = a \bar{z}^2$, where
\begin{align}\label{eq:quad_coeff1}
	a = I_2(N_1,H_1)\left(\widetilde{W}_1-\widetilde{H}_1\right)+G_2(N_1,W_1)\left(\widetilde{N}_1-\gamma \widetilde{W}_1\right),
\end{align}
and $(\widetilde{H}_1, \widetilde{W}_1, \widetilde{N}_1)$ is the left null vector of $J(q_c^2, p_c)$. We choose the convention $\widetilde{H}_1 H_1 + \widetilde{W}_1 W_1 + \widetilde{N}_1 N_1 = 2$, which eliminates an overall factor of 2 in~\eqref{eq:quad_coeff1}.

\section{Turing point calculation and scaling}\label{apdx:wavenumber}
In this appendix, we derive approximations for the critical wave number and the onset parameter value at the upper and lower Turing points, which give scaling relations with respect to the quantity $\delta \equiv I(n_0)/D_h$, where $I(n)$ is the R02 infiltration function, $n_0$ is the uniform vegetated equilibrium biomass, and $D_h$ is the nondimensional surface water diffusion parameter.

We introduce the abbreviations $I_0 \equiv I(n_0)$, $\hat{h}_0 \equiv I'(n_0) h_0$, $\hat{n}_0 \equiv G'(w_0) n_0$, where $(h_0,w_0,n_0)$ is the uniform vegetated steady state given in~\eqref{eq:uniform_veg_ss}. In the following analysis, we will treat $\hat{n}_0$ as the bifurcation parameter, which is justified by the following observations regarding relationships between $\hat{n}_0$, $\hat{h}_0$, and the precipitation parameter $p$:
\begin{enumerate}[I.]
	\item $\hat{n}_0$ is a linear increasing function of $p$. Specifically, $n_0 = (p-p_0)/\gamma \mu$ where $p_0 \equiv \nu \mu/(1-\mu)$, and $\hat{n}_0$ is proportional to $n_0$ by the constant factor $G'(w_0) = (1-\mu)^2$ ($w_0$ depends only on the parameter $\mu$).
	\item $\hat{h}_0$ can be expressed as a function of $\hat{n}_0$, since $\hat{h}_0 \equiv I'(n_0)h_0 = p(n_0) I'(n_0)/I(n_0)$.
	\item We will show that Turing points are confined to regions of the parameter space where $\hat{h}_0  > \gamma \mu$ (see~\eqref{eq:det_zero_eqn2}).
	\item $\hat{h}_0$ is a monotonic decreasing function of $\hat{n}_0$ when $\hat{h}_0 > \gamma \mu$. This follows 	from
	\begin{align*}
		\frac{d\hat{h}_0}{d\hat{n}_0} &= \frac{dn_0}{d\hat{n}_0} \frac{d\hat{h}_0}{dn_0} = \frac{1}{G'(w_0)} \frac{d\hat{h}_0}{dn_0} = \frac{1}{(1-\mu)^2} \left(I''(n_0)h_0 + I'(n_0) \frac{dh_0}{dn_0} \right),
	\end{align*}
	and
	\begin{align*}
		\frac{dh_0}{dn_0} &= \frac{d}{dn_0} \frac{p(n_0)}{I(n_0)} = p'(n_0)\frac{1}{I(n_0)} - I'(n_0) \frac{p}{I(n_0)^2}\\ &= \frac{1}{I(n_0)} \left(\gamma \mu - I'(n_0)h_0  \right) = - \frac{\hat{h}_0 - \gamma \mu}{I(n_0)} < 0 \text{ for } \hat{h}_0 > \gamma \mu.
	\end{align*}
	Since $I(n)$ is a concave-down function (i.e. $I''(n_0) < 0$), it follows that $d\hat{h}_0/d\hat{n}_0 < 0$.
\end{enumerate}
Together, these relationships give a one-to-one correspondence between $\hat{n}_0$, $\hat{h}_0$, and precipitation $p$ in the parameter regime relevant for this analysis. We note that if $\hat{h}_0 < \gamma \mu$ at the lower boundary of the domain, $\hat{n}_0 = 0$, it can never exceed $\gamma \mu$, and no Turing points exist. For Turing points to exist, it must be the case that $\hat{h}_0 > \gamma \mu$ at $\hat{n}_0 = 0$. This gives a necessary condition for a Turing bifurcation in R02:
\begin{align}
	\hat{h}_0(0) = \frac{\nu \mu (1-f)}{f (1 - \mu)} > \gamma \mu.\label{eq:necc_cond}
\end{align}
This condition explains the asymptotic approach of the degeneracy curves in Figures~\ref{fig:weakly_nonlin_results_upper}-\ref{fig:main_result} to $f = \nu/(\nu + \gamma(1-\mu)) = 8/9$ as $\alpha$ decreases or $D_h$ increases.

We write the Jacobian matrix~\eqref{eq:jacobian} as
\begin{align}
	J(q^2,\hat{n}_0) = \begin{pmatrix} -I_0 - D_h q^2 & 0 & - \hat{h}_0  \\ I_0  &  -\gamma \hat{n}_0 - \nu - D_w q^2 & \hat{h}_0 - \gamma \mu \\ 0 & \hat{n}_0  & -q^2 \end{pmatrix}.
\end{align}
The determinant of this matrix is
\begin{align*}
	-D_h D_w q^6 - \left(I_0 D_w + \gamma D_h (\hat{n}_0 + \nu) \right) q^4 - \left(I_0(\gamma \hat{n}_0 + \nu) - D_h(\hat{h}_0 - \gamma \mu)\hat{n}_0 \right) q^2 - I_0 \gamma \mu \hat{n}_0.
\end{align*}
The zero-eigenvalue condition~\eqref{eq:turingconditions1} and the onset condition~\eqref{eq:turingconditions2} are necessary for a Turing point $(q^2_c,\hat{n}_0^c)$, and result in the equations
\begin{align}
	D_w q_c^6 + \left(\delta D_w + \gamma \hat{n}_0^c + \nu \right) q_c^4 + \left( \delta (\gamma \hat{n}_0^c + \nu) - (\hat{h}_0^c -\gamma \mu) \hat{n}_0^c \right) q_c^2 + \delta \gamma \mu \hat{n}_0^c &= 0,\label{eq:det_zero_eqn2}\\
	3 D_w q_c^4 + 2\left(\delta D_w + \gamma \hat{n}_0^c + \nu \right) q_c^2 +\delta (\gamma \hat{n}_0^c + \nu) - (\hat{h}_0^c -\gamma \mu) \hat{n}_0^c &= 0,\label{eq:ddet_zero_eqn}
\end{align}
where
\begin{align*}
	\delta \equiv \frac{I_0}{D_h} \text{ and } \hat{h}_0^c \equiv \hat{h}_0(\hat{n}_0^c).
\end{align*}
We treat $\delta$ as a small parameter by assuming $D_h \gg 1$, noting that $I_0$ is bounded away from zero ($I_0 \in (\alpha f, \alpha)$). The only possible negative term in~\eqref{eq:det_zero_eqn2} and~\eqref{eq:ddet_zero_eqn} comes from the factor $-(\hat{h}_0^c -\gamma \mu)$, and so Turing points are confined to regions of the parameter space where $\hat{h}_0  > \gamma \mu$ (as anticipated above in observation III). We note that these equations depend on $\alpha$ and $D_h$ only through the quantity $\delta \propto \alpha/D_h$. Therefore the Turing point calculation is invariant to fixed ratios of $\alpha/D_h$.

We proceed by seeking approximate Turing point solutions $(q^2_c,\hat{n}_0^c)$ to~\eqref{eq:det_zero_eqn2} and~\eqref{eq:ddet_zero_eqn} in the form of asymptotic expansions in $\delta$. At $\mathcal{O}(1)$ in $\delta$,~\eqref{eq:det_zero_eqn2} and~\eqref{eq:ddet_zero_eqn} are
\begin{align*}
	D_w q^6_c + (\gamma \hat{n}_0^c + \nu) q^4_c - (\hat{h}_0^c - \gamma \mu) \hat{n}_0^c q^2_c &= 0,\\
	3D_w q^4_c + 2(\gamma \hat{n}_0^c + \nu) q^2_c - (\hat{h}_0^c - \gamma \mu) \hat{n}_0^c &=0.
\end{align*}
The only real-valued solution in $q_c^2$ is $q^2_c = 0$, which yields $(\hat{h}_0^c - \gamma \mu) \hat{n}_0^c = 0$. It can be shown that the equation $(\hat{h}_0^c - \gamma \mu) \hat{n}_0^c = 0$ has only two unique solutions, $\hat{n}_0^c = 0$ and $\hat{h}_0^{c} = \gamma \mu$. Thus there are two solutions at this order: (i) $q^2_{\ell} \equiv q^2_c = 0$ and $\hat{n}_0^{\ell} \equiv \hat{n}_0^c = 0$, which corresponds to the lower Turing point, and (ii) $q^2_u \equiv q^2_c = 0$ and $\hat{h}_0^u - \gamma \mu  = 0$ where $\hat{h}_0^u \equiv \hat{h}_0^c = \gamma \mu$, which corresponds to the upper Turing point. We calculate corrections to these solutions, which yield distinct scalings for the wave numbers and onset parameter values with $\delta$ at the different Turing points.

For the lower Turing point solution, we assume the leading-order correction takes the form $(q^2_{\ell}, \hat{n}_0^{\ell}) = (Q^2_{\ell} \delta^{\beta_1}, N_{\ell} \delta^{\beta_2})$, and we seek values of $\beta_1$ and $\beta_2$ which achieve a balance between terms in~\eqref{eq:det_zero_eqn2} and~\eqref{eq:ddet_zero_eqn}. For this correction, equations~\eqref{eq:det_zero_eqn2} and~\eqref{eq:ddet_zero_eqn} become
\begin{multline*}
	 \delta^{3\beta_1} D_w Q_{\ell}^6 + \left(\delta^{2\beta_1 + 1} D_w + \delta^{2\beta_1 + \beta_2} \gamma N_{\ell} + \delta^{2\beta_1} \nu  \right) Q_{\ell}^4 \\
	  + \left( \delta^{\beta_1 +\beta_2 + 1} \gamma N_{\ell} + \delta^{\beta_1 + 1} \nu - \delta^{\beta_1 + \beta_2}(\hat{h}_0^{\ell} -\gamma \mu) N_{\ell} \right) Q_{\ell}^2 + \delta^{\beta_2 + 1} \gamma \mu N_{\ell} = 0,
\end{multline*}\vspace{-2em}
\begin{multline*}
	3\delta^{3\beta_1} D_w Q_{\ell}^4 + 2\left(\delta^{2\beta_1 + 1} D_w + \delta^{2\beta_1 + \beta_2} \gamma N_{\ell} + \delta^{2\beta_1} \nu  \right) Q_{\ell}^2 \\
	+ \delta^{\beta_1 +\beta_2 + 1} \gamma N_{\ell} + \delta^{\beta_1 + 1} \nu - \delta^{\beta_1 + \beta_2}(\hat{h}_0^{\ell} -\gamma \mu) N_{\ell} = 0.
\end{multline*}
We observe that
\begin{itemize}
	\item $\delta^{3\beta_1} D_w Q_{\ell}^6$, $\delta^{2\beta_1 + 1} D_w Q_{\ell}^4$ and $\delta^{2\beta_1 + \beta_2} \gamma N_{\ell} Q_{\ell}^4$ are always higher order terms than $\delta^{2\beta_1} \nu Q_{\ell}^4$.
	\item $- \delta^{\beta_1 + \beta_2}(\hat{h}_0^{\ell} -\gamma \mu) N_{\ell} Q_{\ell}^2$ must appear at leading order for real solutions in $Q^2_{\ell}$.
	\item $\delta^{\beta_1 + \beta_2 + 1} \gamma N_{\ell} Q^2_{\ell}$ is always a higher order term than $-\delta^{\beta_1 + \beta_2}(\hat{h}_0^{\ell} -\gamma \mu) N_{\ell} Q_{\ell}^2$.
	\item $\delta^{\beta_2 + 1} \gamma \mu N_{\ell}$ must appear at leading order for solutions with $Q^2_{\ell}, \ N_{\ell} > 0$.
\end{itemize}
A balance is achieved by $\beta_1 = \beta_2 = 1$. Thus $(q^2_{\ell}, \hat{n}_0^{\ell}) = (Q^2_{\ell} \delta + \mathcal{O}(\delta^2), N_{\ell} \delta + \mathcal{O}(\delta^2))$, and $Q_{\ell}^2$ and $N_{\ell}$ are solutions to the equations at $\mathcal{O}(\delta^2)$:
\begin{equation}
\begin{aligned}
	\nu Q_{\ell}^4 + \left(\nu - (\hat{h}_0^{\ell} -\gamma \mu) N_{\ell} \right) Q_{\ell}^2 + \gamma \mu N_{\ell} &= 0,\\
	2 \nu Q_{\ell}^2 + \nu - (\hat{h}_0^{\ell} -\gamma \mu) N_{\ell} &= 0,\label{eq:lower_coeff_eqn}
\end{aligned}
\end{equation}
where $\hat{h}_0^{\ell} = \hat{h}_0(N_{\ell} \delta) = \nu \mu (1-f)/(f (1-\mu)) + \mathcal{O}(\delta)$. It can be shown that a unique physical solution to this system (i.e. $Q_{\ell}^2, \ N_{\ell} > 0$) exists when the necessary condition~\eqref{eq:necc_cond} is satisfied. See also Dawes~\etal~\citep{Dawes:2015cz} for a comparable scaling of the lower Turing point critical wave number with the nondimensional water diffusion parameter in the model by von Hardenberg~\etal~\citep{vonHardenberg:2001bka}.

For the upper Turing point solution, we assume the leading-order correction takes the form $q^2_u = Q^2_u \delta^{\beta_3}$, $\hat{h}_0^u - \gamma \mu = H_u \delta^{\beta_4}$. Invoking a similar argument as for the lower Turing point correction, we balance terms of order $\delta^{2 \beta_3}$, $\delta^{\beta_3 + \beta_4}$ and $\delta$ and find $\beta_3 = \beta_4 = 1/2$. Thus $(q^2_u, \hat{h}_0^u-\gamma \mu) = (Q^2_u \delta^{1/2} + \mathcal{O}(\delta), H_u \delta^{1/2} + \mathcal{O}(\delta))$, and $Q^2_u$ and $H_u$ are solutions to the leading order equations at $\mathcal{O}(\delta)$:
\begin{equation}
\begin{aligned}
	\left(\gamma \hat{n}_0^u + \nu \right) Q_u^4  - \hat{n}_0^u H_u Q_u^2 + \gamma \mu \hat{n}_0^u &= 0,\\
	2 \left(\gamma \hat{n}_0^u + \nu \right) Q_u^2  - \hat{n}_0^u H_u &= 0,\label{eq:upper_coeff_eqn}
\end{aligned}
\end{equation}
where $\hat{h}_0(\hat{n}_0^u) = \gamma \mu + H_u \delta^{1/2}$. It can be shown that a unique solution to this system, with $Q^2_u > 0$, always exists.

\section{Quadratic coefficient analysis}\label{apdx:quadcoeff_analysis}
In Appendix~\ref{apdx:wavenumber}, we derived scaling relations between $\delta \equiv I(n_0)/D_h$, the critical wave numbers $q_c$ and the Turing point parameter values in terms of $\hat{n}_0 \equiv G'(w_0)n_0$. In this appendix, we use these scaling relations in an analysis of the terms that are important in setting the sign of $a$ when $\delta$ is small.

The leading order scaling behaviour of the quadratic coefficient $a$ is determined by the scaling of the right and left null vectors, $(H_1, W_1, N_1)^{T}$ and $(\widetilde{H}_1, \widetilde{W}_1, \widetilde{N}_1)$ respectively. We will use the relations above to derive the null vector scalings. We recall that $a$ can be written as
\begin{multline}\label{eq:quad_coeff2}
	a = \left( I'(n_0) H_1 N_1 + \frac{1}{2} I''(n_0) h_0 N_1^2 \right)\left(\widetilde{W}_1-\widetilde{H}_1\right) \\+ \left(G'(w_0) W_1 N_1 + \frac{1}{2} G''(w_0) n_0 W_1^2 \right)\left(\widetilde{N}_1-\gamma \widetilde{W}_1\right).
\end{multline}
We obtain the following relations between the right null vector components from the first and third rows of the Jacobian matrix~\eqref{eq:jacobian}:
\begin{align*}
	\left(D_h q_c^2 + I(n_0)\right)H_1 + I'(n_0) h_0 N_1 &= 0,\\
	 G'(w_0)n_0 W_1 - q^2_c N_1 &= 0.
\end{align*}
In addition, we recall the right null vector convention $H_1^2 + W_1^2 + N_1^2 = 1$ with $N_1 > 0$. Together, these equations give
\begin{equation}
\begin{aligned}\label{eq:nullvec}
	H_1 &= -\frac{I'(n_0)h_0}{D_h q^2_c + I(n_0)}N_1, \\
	W_1 &= \frac{q^2_c}{G'(w_0)n_0} N_1, \\
	N_1 &= \left(1+ \frac{q^4_c}{G'(w_0)^2 n_0^2} + \frac{I'(n_0)^2 h_0^2}{(D_h q^2_c + I(n_0))^2} \right)^{-1/2}.
\end{aligned}
\end{equation}
Similarly we obtain the following relations between the left null vector components from the first and second columns of~\eqref{eq:jacobian}:
\begin{align*}
	\left(D_h q_c^2 + I(n_0)\right) \widetilde{H}_1 - I(n_0) \widetilde{W}_1 &= 0, \\
	\left(\nu + D_w q_c^2 + \gamma G'(w_0)n_0 \right) \widetilde{N}_1 + G'(w_0)n_0 \widetilde{W}_1 &= 0.
\end{align*}
Together with the left null vector convention $\widetilde{H}_1 H_1 + \widetilde{W}_1 W_1 + \widetilde{N}_1 N_1 = 2$, we find
\begin{equation}\label{eq:leftnullvec}
	\begin{aligned}
	\widetilde{H_1} &= \frac{I(n_0)}{D_h q^2_c + I(n_0)} \widetilde{W}_1,\\
	\widetilde{W}_1 &= 2 N_1^{-1}\left(\frac{-I(n_0)I'(n_0)h_0}{(D_h q^2_c + I(n_0))^2} + \frac{\nu + (1 + D_w)q^2_c}{G'(w_0)n_0} + \gamma \right)^{-1},\\
	\widetilde{N}_1 &= \left(\frac{\nu + D_w q^2_c}{G'(w_0)n_0} + \gamma\right) \widetilde{W}_1.
	\end{aligned}
\end{equation}

At the lower Turing point, we found that $q^2_{\ell} \equiv q^2_c = Q^2_{\ell} \delta + \mathcal{O}(\delta^2)$ and $G'(w_0)n_0 = N_{\ell} \delta + \mathcal{O}(\delta^2)$, where $Q^2_{\ell}$ and $N_{\ell}$ are given by~\eqref{eq:lower_coeff_eqn}. The right null vector at the lower Turing point,~\eqref{eq:nullvec}, is thus given by
\begin{align*}
	H_1^{\ell} &= -\frac{I'(n_0) h_0}{(1 + Q_{\ell}^2)I(n_0)} N_1^{\ell} + \mathcal{O}(\delta), \\
	W_1^{\ell} &= \frac{Q^2_{\ell}}{N_{\ell}} N_1^{\ell} + \mathcal{O}(\delta), \\
	N_1^{\ell} &= \left(1 + \frac{Q^4_{\ell}}{N^2_{\ell}} + \frac{I'(n_0)^2 h_0^2}{(1+Q_{\ell}^2)^2 I(n_0)^2}\right)^{-1/2} + \mathcal{O}(\delta).
\end{align*}
Similarly the left null vector components are
\begin{align*}
	\widetilde{H}_1^{\ell} &= \delta \frac{2 N_{\ell}}{(1+Q^2_{\ell})\nu} (N_1^{\ell})^{-1} + \mathcal{O}(\delta^2), \\ 	\widetilde{W}_1^{\ell} &= \delta \frac{2 N_{\ell}}{(1+Q_{\ell}^2)\nu} (N_1^{\ell})^{-1} + \mathcal{O}(\delta^2), \\ 	\widetilde{N}_1^{\ell} &= 2 (N_1^{\ell})^{-1} + \mathcal{O}(\delta).
\end{align*}
Substituting these expressions in $a$~\eqref{eq:quad_coeff2} gives
\begin{align*}
	a_{\ell} = \frac{2 Q_{\ell}^2 G'(w_0)}{N_{\ell}} N_1^{\ell} + \mathcal{O}(\delta)
\end{align*}
as the leading order behaviour of the quadratic coefficient at the lower Turing point.
Since $G'(w_0),\ N_{\ell}, \ N_1^{\ell} > 0$, $a_{\ell}$ is positive at leading order, corresponding
to a prediction of spot patterns near the lower Turing point.

At the upper Turing point, given that we are sufficiently far from the degeneracy with the lower Turing point, we found that $q^2_u \equiv q^2_c = Q^2_u \delta^{1/2} + \mathcal{O}(\delta)$, where $Q^2_u$ is determined by~\eqref{eq:upper_coeff_eqn}. Given this, the right null vector components,~\eqref{eq:nullvec}, evaluated at the upper Turing point are
\begin{align*}
	H_1^u &= \frac{-I'(n_0)h_0}{Q^2_u I(n_0)} \delta^{1/2} + \mathcal{O}(\delta), \\
	W_1^u &= \frac{Q^2_u}{n_0 G'(w_0)} \delta^{1/2} + \mathcal{O}(\delta), \\
	N_1^{u} &= 1 + \mathcal{O}(\delta)
\end{align*}
The left null vector components~\eqref{eq:leftnullvec} are
\begin{align*}
	\widetilde{H}_1^u &= \frac{2 G'(w_0)n_0}{Q^2_u \nu + \gamma Q^2_u G'(w_0)n_0} \delta^{1/2} + \mathcal{O}(\delta), \\
	\widetilde{W}_1^u &= \frac{2G'(w_0)n_0}{\nu + \gamma G'(w_0)n_0} + \mathcal{O}(\delta^{1/2}), \\
	\widetilde{N}_1^u &= 2 + \mathcal{O}(\delta^{1/2}).
\end{align*}
Putting these expressions in $a$~\eqref{eq:quad_coeff2} gives
\begin{align*}
	a_u &= \frac{G'(w_0) I''(n_0) h_0 n_0}{\nu + \gamma G'(w_0)n_0} + \mathcal{O}(\delta^{1/2}).
\end{align*}
Since $I(n_0)$ is a concave-down function, $I''(n_0) < 0$, and thus $a_u$ is negative at leading order at the upper Turing point. This corresponds to a prediction of gap patterns near the upper Turing point.


\bibliography{library}

\begin{thebibliography}{10}

\bibitem{vonHardenberg:2001bka}
von Hardenberg J, Meron E, Shachak M, Zarmi Y.
\newblock {Diversity of Vegetation Patterns and Desertification}.
\newblock Phys Rev Lett. 2001 Oct;87(19):198101.

\bibitem{Bel:2012jqa}
Bel G, Hagberg A, Meron E.
\newblock {Gradual regime shifts in spatially extended ecosystems}.
\newblock Theor Ecol. 2012 Jan;5(4):591--604.

\bibitem{Dakos:2011dpa}
Dakos V, K{\'e}fi S, Rietkerk M, van Nes EH, Scheffer M.
\newblock {Slowing Down in Spatially Patterned Ecosystems at the Brink of
  Collapse}.
\newblock Am Nat. 2011 Jun;177(6):E153--E166.

\bibitem{Gilad:2004bp}
Gilad E, von Hardenberg J, Provenzale A, Shachak M, Meron E.
\newblock {Ecosystem engineers: from pattern formation to habitat creation}.
\newblock Phys Rev Lett. 2004;93(9):098105.

\bibitem{Gowda:2014cd}
Gowda K, Riecke H, Silber MC.
\newblock {Transitions between patterned states in vegetation models for
  semiarid ecosystems}.
\newblock Phys Rev E. 2014;89(2):022701.

\bibitem{Guttal:2007ixa}
Guttal V, Jayaprakash C.
\newblock {Self-organization and productivity in semi-arid ecosystems:
  Implications of seasonality in rainfall}.
\newblock J Theor Biol. 2007 Oct;248(3):490--500.

\bibitem{Kefi:2010gca}
K{\'e}fi S, Eppinga MB, de~Ruiter PC, Rietkerk M.
\newblock {Bistability and regular spatial patterns in arid ecosystems}.
\newblock Theor Ecol. 2010 Jan;3(4):257--269.

\bibitem{LeJeune:2002eoa}
LeJeune O, Tlidi M, Couteron P.
\newblock {Localized vegetation patches: A self-organized response to resource
  scarcity}.
\newblock Phys Rev E. 2002 Jul;66(1):010901.

\bibitem{Meron:2004dra}
Meron E, Gilad E, von Hardenberg J, Shachak M, Zarmi Y.
\newblock {Vegetation patterns along a rainfall gradient}.
\newblock Chaos Solitons Fractals. 2004 Jan;19(2):367--376.

\bibitem{Sherratt:2015cq}
Sherratt JA.
\newblock {Using wavelength and slope to infer the historical origin of
  semiarid vegetation bands}.
\newblock Proc Natl Acad Sci USA. 2015 Mar;p. 201420171.

\bibitem{Zelnik:2013iv}
Zelnik YR, Kinast S, Yizhaq H.
\newblock {Regime shifts in models of dryland vegetation}.
\newblock Phil Trans R Soc A. 2013;.

\bibitem{vanderStelt:2013wy}
van~der Stelt S, Doelman A, Hek G, Rademacher JD.
\newblock {Rise and fall of periodic patterns for a generalized
  Klausmeier--Gray--Scott model}.
\newblock J Nonlinear Sci. 2013;23(1):39--95.

\bibitem{Siteur:2014jm}
Siteur K, Siero E, Eppinga MB, Rademacher JD.
\newblock {Beyond Turing: The response of patterned ecosystems to environmental
  change}.
\newblock Ecol Complex. 2014;.

\bibitem{Rietkerk:2002ufa}
Rietkerk M, Boerlijst MC, van Langevelde F, HilleRisLambers R, van~de Koppel J,
  Kumar L, et~al.
\newblock {Self-organization of vegetation in arid ecosystems}.
\newblock Am Nat. 2002;160(4):524--530.

\bibitem{LeJeune:2004bma}
LeJeune O, Tlidi M, Lefever R.
\newblock {Vegetation spots and stripes: Dissipative structures in arid
  landscapes}.
\newblock Int J Quant Chem. 2004;98(2):261--271.

\bibitem{Rietkerk:2004vqa}
Rietkerk M, Dekker SC, de~Ruiter PC, van~de Koppel J.
\newblock {Self-organized patchiness and catastrophic shifts in ecosystems}.
\newblock Science. 2004;305(5692):1926--1929.

\bibitem{Scheffer:2009wj}
Scheffer M, Bascompte J, Brock WA, Brovkin V, Carpenter SR, Dakos V, et~al.
\newblock {Early-warning signals for critical transitions}.
\newblock Nature. 2009;461(7260):53--59.

\bibitem{Turing:1952vn}
Turing AM.
\newblock {The Chemical Basis of Morphogenesis}.
\newblock Phil Trans R Soc B. 1952 Aug;237(641):37--72.

\bibitem{Barbier:2006jwa}
Barbier N, Couteron P, LeJoly J, Deblauwe V, LeJeune O.
\newblock {Self-organized vegetation patterning as a fingerprint of climate and
  human impact on semi‐arid ecosystems}.
\newblock J Ecol. 2006;94(3):537--547.

\bibitem{Deblauwe:2011ee}
Deblauwe V, Couteron P, LeJeune O, Bogaert J, Barbier N.
\newblock {Environmental modulation of self-organized periodic vegetation
  patterns in Sudan}.
\newblock Ecography. 2011 Jul;34(6):990--1001.

\bibitem{Klausmeier:1999woa}
Klausmeier CA.
\newblock {Regular and irregular patterns in semiarid vegetation}.
\newblock Science. 1999;284(5421):1826--1828.

\bibitem{Gilad:2007eh}
Gilad E, von Hardenberg J, Provenzale A, Shachak M, Meron E.
\newblock {A mathematical model of plants as ecosystem engineers}.
\newblock J Theor Biol. 2007 Feb;244(4):680--691.

\bibitem{Hoyle:2006ur}
Hoyle R.
\newblock {Pattern Formation: An Introduction to Methods}.
\newblock 1st ed. Cambridge University Press; 2006.

\bibitem{Cross:2009vp}
Cross M, Greenside H.
\newblock {Pattern Formation and Dynamics in Nonequilibrium Systems}.
\newblock 1st ed. Cambridge University Press; 2009.

\bibitem{Judd:2000wma}
Judd SL, Silber MC.
\newblock {Simple and superlattice Turing patterns in reaction--diffusion
  systems: bifurcation, bistability, and parameter collapse}.
\newblock Physica D. 2000;136(1):45--65.

\bibitem{Golubitsky:2012ud}
Golubitsky M, Stewart I, Schaeffer DG.
\newblock {Singularities and Groups in Bifurcation Theory}.
\newblock Springer Verlag; 2012.

\bibitem{Wiggins:2003wd}
Wiggins S.
\newblock {Introduction to Applied Nonlinear Dynamical Systems and Chaos}.
\newblock Springer; 2003.

\bibitem{Cox:2002cga}
Cox SM, Matthews PC.
\newblock {Exponential Time Differencing for Stiff Systems}.
\newblock J Comput Phys. 2002 Mar;176(2):430--455.

\bibitem{Kassam:2005jva}
Kassam AK, Trefethen LN.
\newblock {Fourth-Order Time-Stepping for Stiff PDEs}.
\newblock SIAM J Sci Comput. 2005 Jan;26(4):1214--1233.

\bibitem{Kassam:2003ws}
Kassam AK.
\newblock {Solving reaction-diffusion equations 10 times faster}.
\newblock University of Oxford; 2003.

\bibitem{Siero:2015ht}
Siero E, Doelman A, Eppinga MB, Rademacher JD, Rietkerk M, Siteur K.
\newblock {Striped pattern selection by advective reaction-diffusion systems:
  Resilience of banded vegetation on slopes}.
\newblock Chaos. 2015 Mar;25(3):036411.

\bibitem{Kolokolnikov:2006hi}
Kolokolnikov T, Ward MJ, Wei J.
\newblock {Zigzag and breakup instabilities of stripes and rings in the
  two-dimensional Gray-Scott model}.
\newblock Stud Appl Math. 2006;116(1):35--95.

\bibitem{Vanag:2007dt}
Vanag VK.
\newblock {Waves and patterns in reaction--diffusion systems.
  Belousov--Zhabotinsky reaction in water-in-oil microemulsions}.
\newblock Phys Usp. 2007 Oct;47(9):923--941.

\bibitem{Malchow:2004ema}
Malchow H, Hilker FM, Petrovskii SV, Brauer K.
\newblock {Oscillations and waves in a virally infected plankton system: Part
  I: The lysogenic stage}.
\newblock Ecol Complex. 2004;1(3):211--223.

\bibitem{Deblauwe:2008if}
Deblauwe V, Barbier N, Couteron P, LeJeune O, Bogaert J.
\newblock {The global biogeography of semi-arid periodic vegetation patterns}.
\newblock Global Ecol Biogeogr. 2008 Nov;17(6):715--723.

\bibitem{Zelnik:2015kf}
Zelnik YR, Meron E, Bel G.
\newblock {Gradual regime shifts in fairy circles}.
\newblock Proc Natl Acad Sci USA. 2015 Sep;p. 201504289.

\bibitem{Hillel:2013tn}
Hillel D.
\newblock {Introduction to Soil Physics}.
\newblock Academic Press; 2013.

\bibitem{Getzin:2014kt}
Getzin S, Wiegand K, Wiegand T, Yizhaq H, von Hardenberg J, Meron E.
\newblock {Adopting a spatially explicit perspective to study the mysterious
  fairy circles of Namibia}.
\newblock Ecography. 2014 May;38(1):1--11.

\bibitem{Iams:2015rk}
Iams S, Gowda K, Silber MC.
\newblock {Comparing reaction-diffusion models of vegetation patterns in
  semiarid ecosystems}.
\newblock Manuscript in preparation. 2016;.

\bibitem{Dawes:2015cz}
Dawes J, Williams J.
\newblock {Localised pattern formation in a model for dryland vegetation}.
\newblock J Math Biol. 2015 Oct;p. 1--28.

\end{thebibliography}


\begin{thebibliography}{1}

\bibitem{Rietkerk:2002ufa}
Rietkerk M, Boerlijst MC, van Langevelde F, HilleRisLambers R, van~de Koppel J,
  Kumar L, et~al.
\newblock {Self-organization of vegetation in arid ecosystems}.
\newblock Am Nat. 2002;160(4):524--530.

\bibitem{Cox:2002cga}
Cox SM, Matthews PC.
\newblock {Exponential Time Differencing for Stiff Systems}.
\newblock J Comput Phys. 2002 Mar;176(2):430--455.

\bibitem{Kassam:2005jva}
Kassam AK, Trefethen LN.
\newblock {Fourth-Order Time-Stepping for Stiff PDEs}.
\newblock SIAM J Sci Comput. 2005 Jan;26(4):1214--1233.

\bibitem{Kassam:2003ws}
Kassam AK.
\newblock {Solving reaction-diffusion equations 10 times faster}.
\newblock University of Oxford; 2003.

\end{thebibliography}
\bibliographystyle{vancouver}

\end{document}


\title{Supplementary information}
\subtitle{Assessing the robustness of spatial pattern sequences in a dryland vegetation model}
\author{Karna Gowda\thanks{Department of Engineering Sciences and Applied Mathematics, Northwestern University, Evanston, IL 60208, USA}, Yuxin Chen\footnotemark[1], Sarah Iams\thanks{Paulson School of Engineering and Applied Sciences, Harvard University, Cambridge, MA 02138, USA}, Mary Silber\thanks{Department of Statistics, The University of Chicago, Chicago, IL 60637, USA}}
\date{}

\maketitle

\section{Numerical simulation procedure}\label{sec:exploration_procedure}
We numerically solved the Rietkerk~\etal~\citep{Rietkerk:2002ufa} (R02) model using the exponential time differencing Runge-Kutta 4 (ETDRK4) scheme~\citep{Cox:2002cga,Kassam:2005jva} modified for 2D systems~\citep{Kassam:2003ws}. ETDRK4 achieves pseudospectral accuracy in space and fourth-order accuracy in time. This scheme alleviates issues of stiffness often associated with reaction-diffusion systems~\citep{Kassam:2003ws}, allowing R02 to be simulated efficiently. Simulations were run using a time step near the scheme's empirically derived stability limit, since our primary concern is with qualitative aspects of patterned solutions. This scheme was implemented in MATLAB and C.

For each set of parameter values marked by a letter in Figure 7, we simulated R02 over a range of precipitation values using a procedure described in detail here. Simulations are initialised with spatially random initial conditions with fields taking uniformly distributed values in $[1,1.5]$. The simulation begins just below the upper Turing point, at $p^0 = p_{\ell} + 0.95(p_u - p_{\ell})$. Then the following loop is run to identify the upper bound of pattern stability beginning with $k = 0$:
\begin{enumerate}[1.]
	\item R02 is solved using ETDRK4 until either a steady state stop condition or a uniformity stop condition is reached. The end state of the simulation, $(h,w,n)^k$ at $p^k$, is saved.
	\item If the steady state stop condition is reached, precipitation is incremented upward by a small value $\Delta p$, so that $p^{k+1} = p^k + \Delta p$. The procedure returns to step 1, using the saved end state $(h,w,n)^k$ as the initial condition for a new simulation. If the spatial uniformity stop condition is reached, the loop ends.
\end{enumerate}
The steady state stop condition occurs when either (a) the root mean square difference between the current biomass state and the state 400 time units earlier drops below $10^{-4}$, or (b) $t = 2 \times 10^5$. The uniformity stop condition occurs when the root mean square difference between the current biomass state and the mean value of that state drops below $10^{-4}$. We infer the upper bound of pattern stability, $p_{u+} = p^{N}$, to be the value of precipitation where the uniformity condition is reached.

After the first loop terminates at $p^N$ and the first uniformity condition is reached, precipitation is decremented by $2\Delta p$ so that $p^{N+1} = p^N - 2 \Delta p$. The previous patterned (i.e. non-uniform) end state, $(h,w,n)^{N-1}$, is used as the initial condition for a new simulation. We run the following loop to identify pattern morphologies that occur as precipitation decreases, starting with $k = N+1$:
\begin{enumerate}[1.]
	\item R02 is solved using ETDRK4 until a stop condition is reached. The end state of the simulation, $(h,w,n)^k$, is saved.
	\item If a steady state stop condition is reached, precipitation is decremented by $\Delta p$ so that $p^{k+1} = p^k - \Delta p$. The procedure returns to step 1, using the saved end state $(h,w,n)^k$ as the initial condition for a new simulation. If the spatial uniformity stop condition is reached, the loop ends.
\end{enumerate}
When this loop ends at $p^{N+M}$, we infer this point to be the lower bound of pattern stability, $p_{\ell-} = p^{N+M}$.

We increment precipitation upward in one final loop to identify any potential hysteresis effects. Precipitation is first incremented by $2\Delta p$ so that $p^{N+M+1} = p^{N+M} + 2\Delta p$. The previous patterned (non-uniform) end state $(h,w,n)^{N+M-1}$ is used as the initial condition for a new simulation. Then the following loop is run, starting with $k = N+M+1$:
\begin{enumerate}[1.]
	\item R02 is solved using ETDRK4 until a stop condition is reached. The end state of the simulation, $(h,w,n)^{k}$, is saved.
	\item If a steady state stop condition is reached, precipitation is incremented by $\Delta p$ so that $p^{k+1} = p^k + \Delta p$. The procedure returns to step 1, using the saved end state as the initial condition for a new simulation. If the spatial uniformity stop condition is reached, the loop ends.
\end{enumerate}
The end result of this procedure is a series of saved end states at a range of precipitation values over the interval $p \in (p_{\ell-}, p_{u+})$.

The precipitation iteration step size $\Delta p$ was chosen based on the distance between the upper and lower Turing points, $p_u - p_{\ell}$, so that 30-100 end states were saved per parameter set. The value of $\Delta p$ used ranged between $0.0025$ and $0.01$. For most simulations, a time step size of $\Delta t = 0.4$ was used, which was near the empirical stability limit of the scheme for most parameter sets. On certain parameter sets, values of $\Delta t$ as small as $0.01$ were needed for numerical stability. We tested for time step size errors using a particular parameter set, with $\Delta t = 0.4,\ 0.2,$ and $0.1$, and the end states were found to be qualitatively indistinguishable. A preliminary set of simulations was run on all parameter sets to identify a simulation domain size for each that permitted at least 7 wavelengths of patterns. Square $L \times L$ domains with $L = 400,\ 800,$ and $1600$ were used. Corresponding $N \times N$ grid sizes were used, with $N = 64,\ 128,$ and $256$ respectively.

\section{Investigating the ``stripes $\to$ spots'' sequences}
Our bifurcation analysis indicates that gap solutions to the amplitude equations should be stable near the upper Turing point (region B of Figure 5). When $a$ is sufficiently small, as it may be for the ``stripes $\to$ spots'' observations, the gaps solution is stable for only a relatively small interval of the precipitation parameter compared to the stripes branch. This is illustrated schematically in bifurcation diagram B in Figure 7. Because our numerical procedure increments precipitation in discrete steps of fixed size, the gaps branch may be bypassed. Also, since gaps are stable only very near to the Turing point for these parameter sets, gaps may only be stable at the critical wavelength and may not appear in a domain of arbitrary size.

To test whether gaps can exist stably for parameter sets where ``stripes $\to$ spots'' transitions are observed, we seeded numerical simulations at $p = p_u$ with a hexagonal lattice gaps pattern at the critical wavelength. This initial condition was perturbed by spatially random noise drawn from a uniform distribution on the interval $[0,0.1]$. An aspect ratio of $1:\sqrt{3}$ and a domain size that permits an integer multiple of the critical wavelength were used. In all simulations, gap patterns persisted as steady states through $t = 1 \times 10^6$ and were assessed to be  stable in these instances.

\section{Investigating the ``stripes $\to$ spirals'' sequence}
We observed time-varying spiral wave patterns in one instance of the numerical simulations, at $f = 0.1$ and $\log_{10}(D_h) = 0.5$. States from this simulation are shown in Supplementary Figure~\ref{fig:spirals}. We verified that the observation of spirals is robust to increased spatial and temporal resolution in the numerical scheme. We ran additional simulations at nearby parameter values to determine whether spirals are confined to a region of the parameter space. Holding $f=0.1$ fixed, spirals appear in simulations at $\log_{10}(D_h) = 0.6$ but not $\log_{10}(D_h) = 0.7$. Holding $\log_{10}(D_h) = 0.5$ fixed, spirals appear at $f=0.16$ but not at $f=0.18$. We conclude that spirals are confined to smaller values of $D_h$ than are typically considered ecologically applicable.

\begin{suppfigure}
	\centering \includegraphics[width=0.6\columnwidth]{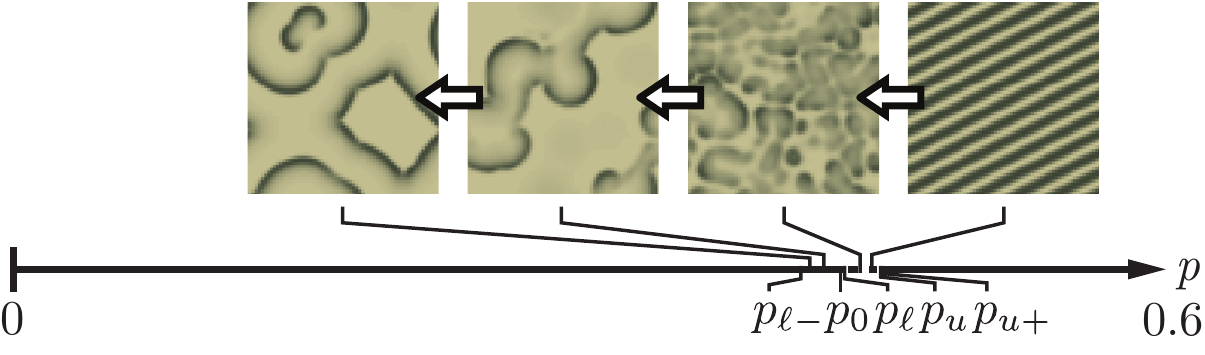} \caption{\small Spiral wave patterns are observed in simulations at $f = 0.1$ and $\log_{10}(D_h) = 0.5$. An ordered stripes state transitions directly to spirals at $p = 0.414$.} \label{fig:spirals}
\end{suppfigure}

\bibliography{library}
\bibliographystyle{vancouver}